\documentclass[12pt,a4j]{article}
\usepackage[dvips]{graphicx}
\usepackage{amsmath}
\usepackage{amssymb}
\usepackage{enumerate}

\begin{document}
\baselineskip=7mm
\centerline{\bf  Smooth and singular multisoliton solutions of a modified Camassa-Holm }\par
\centerline{\bf equation with cubic nonlinearity and linear dispersion}\par
\bigskip
\centerline{Yoshimasa Matsuno\footnote{{\it E-mail address}: matsuno@yamaguchi-u.ac.jp}}\par
\centerline{\it Division of Applied Mathematical Science, Graduate School of Science and Engineering} \par
\centerline{\it Yamaguchi University, Ube, Yamaguchi 755-8611, Japan} \par
\bigskip
\bigskip
\bigskip
\noindent {\bf Abstract} \par
\noindent\ We develop a direct method for solving a modified Camassa-Holm  equation with cubic nonlinearity
and linear dispersion under the rapidly decreasing boundary condition. 
We obtain  a compact parametric representation for the multisoliton solutions and investigate their properties. We show that
the introduction of a linear dispersive term  exhibits various
new features in the structure of solutions. In particular, we find the smooth solitons whose characteristics are different from those of the Camassa-Holm equation, as well as
the novel types of singular solitons.
A remarkable feature of  the soliton solutions is that the underlying structure of the associated tau-functions is the same as that of a
model equation for shallow-water waves introduced by Ablowitz {\it et al} (1974 {\it Stud. Appl. Math.} {\bf 53} 249-315). 
Finally, we demonstrate that  the short-wave limit of the soliton solutions recovers the soliton solutions
of the short pulse equation which describes the propagation of ultra-short optical pulses in nonlinear media.
\par

\newpage
\leftline{\bf  1. Introduction} \par
\bigskip
\noindent In this paper, we consider the following  modified Camassa-Holm (mCH) equation with cubic nonlinearity and linear dispersion
$$ m_t+2\kappa^2u_x+[m(u^2-u_x^2)]_x=0,\quad  m=u-u_{xx}, \eqno(1.1)$$
where $u=u(x,t)$ is a real-valued function of  time $t$ and a spatial variable $x$, and the subscripts $x$ and $t$ appended to $m$ and $u$ denote partial differentiation.
The positive parameter $\kappa$ characterizes the magnitude of the linear dispersion.
The mCH equation was introduced independently by several researchers using a novel  procedure that generates new integrable systems 
from  known integrable bi-hamiltonian systems [1-3]. Actually, the method yields the CH equation [4] when applied to the  Korteweg-de Vries (KdV) equation
whereas it yields the mCH equation when applied to the modified KdV equation. 
In a physical context, it was derived from the two-dimensional Euler equation by using a singular perturbation method in which
 the variable $u$ represents the  velocity of fluid [5]. It also arises from an intrinsic invariant planar curve flow in Euclidian geometry [6].
 The mCH equation admits a Lax pair representation and hence, in principle, it may be solved by means of the inverse scattering transform (IST) method [7]. 
 \par
 The dispersionless version of the mCH equation (equation (1.1) with $\kappa=0$) has attracted  much attention  and has been studied extensively in recent years. 
  There exists a variety of solutions depending on the boundary conditions.  Specifically, under the vanishing boundary condition,
 it exhibits 
the usual peakons [6]  whereas  under the nonvanishing boundary condition, it supports smooth bright soliton solutions [8, 9]. 
See also [10]  for the Lie algebraic approach for constructing solutions. 
 A stability analysis reveals that the single peakon is stable for small perturbations in the energy space [11]. If, on the other hand,  one includes the linear
 dispersion as shown in equation (1.1), then various new features appear in the structure of solutions. 
 In particular, it will admit smooth soliton solutions which vanish at infinity, unlike the dispersionless mCH equation for which the nonexistence result 
 for smooth  traveling wave solutions  has been established under the vanishing boundary condition [5, 12].
 Some qualitative results for the Cauchy problem of equation (1.1) were also reported in a later work; these are mainly concerned with the local well-posedness for the strong solutions
 and the blow-up phenomena [12].
 The scattering problem for equation (1.1) was investigated recently by means of the IST and the time evolution of the scattering data was derived [13]. However, the
 resolution of the  inverse problem  remains open. \par
  Another interesting aspect of the mCH equation is that it
 reduces to the short pulse (SP) equation 
 $$v_{tx}=2\kappa^2 v+{1\over 3}(v^3)_{xx},\qquad v=v(x, t), \eqno(1.2)$$
 in the short-wave limit [6].  Equation (1.2) describes the propagation of ultra-short optical pulses in nonlinear media [14].
 Soliton and periodic solutions to the equation are known and their properties  have been explored in detail [15-17].
   \par
  The main purpose of this paper is to construct soliton solutions of the mCH equation which decay
rapidly at infinity and investigate their properties.
We will show that it admits smooth soliton solutions like the bright solitons and breathers as well as the singular solitons. 
 A direct method is employed to obtain solutions which mimics the construction of the soliton solutions of 
 the generalized sine-Gordon (sG) equation [18, 19], the dispersionless  mCH equation [9] and the Novikov equation [20]. \par
 This paper is organized as follows. In section 2, we transform the mCH equation to a system of nonlinear partial differential equations (PDEs) by a
 reciprocal transformation. We then apply a dependent variable transformation to reduce this system   to a system of bilinear equations for 
 the tau-functions. Subsequently, we analyze the latter system by means of  the bilinear transformation method and present a
 compact parametric representation  for the $N$-soliton solution of the mCH equation, where $N$ is an arbitrary positive integer.
Remarkably, we find that the underlying structure of the tau-functions constituting the $N$-soliton solution  is essentially  the same as that of the
$N$-soliton solution of a model equation for shallow-water waves introduced by Ablowitz {\it et al} [21].
We remark that the same statement is true for the tau-functions of the $N$-soliton solutions of the CH [22-25] and dispersionless mCH [9] equations.
In section 3, we investigate the properties of the solutions focusing  on the one- and two-soliton solutions. 
First, we show that the  smooth soliton exists  if the amplitude parameter of the soliton satisfies 
a certain condition.  Furthermore, we obtain  two types of the singular solitons. One has a symmetric profile 
and the other  an antisymmetric profile. 
We analyze the limiting profiles of these singular solitons when the dispersion parameter $\kappa$ tends to zero and find that they do not
recover the peakons constructed in [6].
Subsequently, we perform the asymptotic analysis of the smooth two-soliton solution and confirm its
solitonic behavior. We find that the formula for the phase shift coincides precisely with that of the two-soliton solution of the CH equation.
We also describe briefly the interaction of a smooth soliton and a singular symmetric soliton. 
We then construct the breather solution by specifying the complex conjugate pair for the amplitude parameters characterizing the smooth two-soliton solution.
Finally, we  address the general $N$-soliton solution and present the
formula for the phase shift. In section 4, we demonstrate that the $N$-soliton solution reduces to the $N$-soliton solution of the SP equation
(1.2) under an appropriate limiting procedure. Section 5 is devoted to some concluding remarks. In the appendix, we prove the bilinear identities for the tau-functions 
associated with the $N$-soliton solution  by means of mathematical induction. \par
\bigskip
\noindent{\bf 2. Exact method of solution}\par
\bigskip
\noindent In this section, we develop a systematic procedure for solving the mCH equation by means of the 
bilinear transformation method [26, 27]. 
Specifically, we seek soliton solutions which decay rapidly at infinity. 
We show that the system of
bilinear equations deduced from the mCH equation is closely related to that of the generalized sG equation [18, 19].
This fact helps us to construct soliton solutions of the mCH equation. \par
\bigskip
\noindent{\it 2.1. Reciprocal transformation}\par
\medskip
\noindent First, we rewrite the mCH in the  form of the  conservation law
$$ r_t+[r(u^2-u_x^2)]_x=0, \qquad r=\sqrt{m^2+\kappa^2}. \eqno(2.1)$$
This enables us to introduce the coordinate transformation 
$(x ,t) \rightarrow (y, \tau)$ by
$$dy=rdx-r(u^2-u_x^2)dt, \qquad d\tau=dt. \eqno(2.2a)$$ 
 The $x$ and $t$ derivatives transform  as
 $${\partial\over\partial x}=r{\partial\over\partial y}, \qquad {\partial\over\partial t}
 ={\partial\over\partial \tau}-r(u^2-u_x^2){\partial\over\partial y}. \eqno(2.2b)$$
 Applying the transformation (2.2) to equation (2.1), we can recast it into the form
$$r_\tau+2r^2mu_y=0. \eqno(2.3)$$
It then follows from (2.2b) that the variable $x=x(y, \tau)$ obeys a system of linear PDEs
$$x_y={1\over r}, \eqno(2.4a)$$
$$ x_\tau=u^2-r^2u_y^2. \eqno(2.4b)$$
The system of equations (2.4) is integrable since its compatibility condition $x_{\tau y}=x_{y\tau}$ is assured by virtue of
equation (2.3). \par
If we define the new angular variable $\phi=\phi(y, \tau)$ by
$$m=\kappa\,\tan \,\phi,\eqno(2.5)$$
then $r$ from (2.1) can be represented in terms of $\phi$ as
$$r={\kappa\over \cos\,\phi}, \eqno(2.6)$$
with $-\pi/2<\phi<\pi/2$. 
Substitution of  (2.5) and (2.6) into equation (2.3) gives
$$u_y+{1\over 2\kappa^2}(\sin\,\phi)_\tau=0. \eqno(2.7)$$
\par
Next, we  express $u$  in the form $u=m+r^2u_{yy}+rr_yu_y$ and 
rewrite this expression in terms of $u$ and $\phi$ with the aid of  (2.5)-(2.7) and obtain the equation
$$\phi_{\tau y}+2u\,\cos\,\phi-2\kappa\,\sin\,\phi=0. \eqno(2.8)$$
Using the relation $ru_y=-\phi_\tau/(2\kappa)$ which follows from (2.6) and (2.7), 
the linear system (2.4) can be put into the form
$$x_y={1\over\kappa}\cos\,\phi, \eqno(2.9a)$$
$$ x_\tau=u^2-{1\over 4\kappa^2}\,\phi_\tau^2. \eqno(2.9b)$$
Note that the $x$ derivative of $u$ is  expressed simply  as
$$u_x=-{1\over 2\kappa}\,\phi_\tau. \eqno(2.10)$$
\par
The system of nonlinear PDEs (2.7) and (2.8) for $u$ and $\phi$  is the starting point in the following analysis. 
Actually, the procedure for constructing solutions consists of two steps. First, solve this system
under the boundary conditions $u\rightarrow 0$ and $\phi\rightarrow 0$ as $|y|\rightarrow \infty$
to obtain $u$ and $\phi$ as functions of $y$ and $\tau$. Subsequently, integrate equation (2.9a) to 
give the mapping from $y$ to $x$
$$x={y\over\kappa}+{1\over\kappa}\int^y_{-\infty}(\cos\,\phi-1)\,dy+d, \eqno(2.11)$$
where $d$ is an integration constant depending generally on $\tau$. If we differentiate (2.11) by $\tau$
and use  (2.9b) as well as (2.7) and (2.8), we  confirm that $d^\prime(\tau)=0$ and so this constant is indeed independent of $\tau$. 
Last, performing the integration with respect to $y$ in (2.11),  we obtain a parametric representation for the solution
of the form $u=u(y, \tau), x=x(y, \tau)$. 
\par
\bigskip
\leftline{\it 2.2. Bilinearization} \par
\noindent Here, we perform the procedure for constructing soliton solutions as described in section 2.1.  First, we bilinearize the system of PDEs
(2.7) and (2.8) by introducing a dependent variable transformation.  To this end, we seek solutions of the form
$$u={1\over 2{\rm i}\kappa}\left({\rm ln}\,{\widetilde F\over F}\right)_\tau,\qquad F=F(y, \tau),\qquad \widetilde F=\widetilde F(y, \tau), \eqno(2.12a)$$
$$\phi={\rm i}\,{\rm ln}\,{\widetilde G\over G},\qquad G=G(y, \tau), \qquad \widetilde G=\widetilde G(y, \tau), \eqno(2.12b)$$
subjected to the boundary conditions $F, \widetilde F, G$ and $\widetilde G \rightarrow 1$ as $y\rightarrow -\infty$, where
$F, \widetilde F, G$ and $\widetilde G$  are tau-functions. 
Substituting (2.12) into (2.7) and integrating the resultant expression by $\tau$ under the boundary conditions specified above, we obtain
$${1\over 2{\rm i}\kappa}\left({\rm ln}\,{\widetilde F\over F}\right)_y+{1\over 4{\rm i}\kappa^2}\left({G\over \widetilde G}-{\widetilde G\over G}\right)=0.\eqno(2.13)$$
If we impose an auxiliary condition 
$$\widetilde FF=\widetilde GG, \eqno(2.14)$$
for the tau-functions, then we can transform (2.13) into the bilinear equation
$$D_y\widetilde F\cdot F+{1\over 2\kappa}(G^2-{\widetilde G}^2)=0, \eqno(2.15)$$
where the bilinear operators  are defined by
$$D_y^mD_\tau ^nF\cdot G=\left(\partial_y-\partial_{y^\prime} \right)^m\left(\partial_\tau-\partial_{\tau^\prime} \right)^n
F(y,\tau)G(y^\prime,\tau^\prime)|_{y^\prime=y,\,\tau^\prime=\tau}, \qquad (m, n=0, 1, 2, ...).   \eqno(2.16)$$
\par
To solve equation (2.8), we use the following relations which stem from (2.12) and the definition of the bilinear operators:
$$\phi_{\tau y}=-{{\rm i}\over 2G^2}D_\tau D_yG\cdot G+{{\rm i}\over 2{\widetilde G}^2}D_\tau D_y\widetilde G\cdot \widetilde G, \eqno(2.17a)$$
$$u\,\cos\,\phi={1\over 4{\rm i}\kappa}{(G^2+{\widetilde G}^2)D_\tau \widetilde F\cdot F\over \widetilde FF\widetilde GG},\eqno(2.17b)$$
$$\kappa\,\sin\,\phi={\kappa\over 2{\rm i}}{G^2-{\widetilde G}^2\over \widetilde GG}. \eqno(2.17c)$$
Substituting (2.17) into (2.8) and using  (2.14), we can recast (2.8) into the form
$${\widetilde G}^2\left(D_\tau D_yG\cdot G+{1\over \kappa}D_\tau \widetilde F \cdot F+2\kappa \widetilde GG\right)
=G^2\left(D_\tau D_y\widetilde G\cdot \widetilde G +{1\over \kappa}D_\tau F\cdot \widetilde F+2\kappa G\widetilde G\right). \eqno(2.18)$$
We  decouple the above equation as
$$D_\tau D_yG\cdot G+{1\over \kappa}D_\tau \widetilde F\cdot F+2\kappa \widetilde GG=\mu G^2, \eqno(2.19a)$$
$$D_\tau D_y\widetilde G\cdot \widetilde G +{1\over \kappa}D_\tau F\cdot \widetilde F+2\kappa G\widetilde G=\mu {\widetilde G}^2, \eqno(2.19b)$$
by introducing  a parameter  $\mu$ which depends generally on $y$ and $\tau$. To determine $\mu$, we divide (2.19$a$) by $G^2$, take the limit $y\rightarrow -\infty$
and use the boundary conditions for $F, \widetilde F, G$ and $\widetilde G$. We then find that $\mu=2\kappa$ which, substituted in (2.19), gives
$$D_\tau D_yG\cdot G+{1\over \kappa}D_\tau \widetilde F\cdot F+2\kappa (\widetilde G -G)G=0, \eqno(2.20a)$$
$$D_\tau D_y\widetilde G\cdot \widetilde G +{1\over \kappa}D_\tau F\cdot \tilde F+2\kappa (G-\widetilde G)\widetilde G=0. \eqno(2.20b)$$
\par
Thus, the problem under consideration has been reduced to solving the system of bilinear equations (2.15) and (2.20)
for $F, \widetilde F, G$ and $\widetilde G$ subject to condition (2.14).  
 Fortunately,  we have encountered a similar problem in the analysis of
the generalized sG equation. Specifically, we recall that the bilinear equation (2.15) is essentially the same as 
 (2.23$a$) of [18] if the asterisk is replaced by the tilde.  Bearing this in mind, we put
$$F=fg,\qquad \widetilde F=\tilde f\tilde g,\qquad G=f\tilde g, \qquad \tilde G=\tilde fg, \eqno(2.21)$$
where $f, \tilde f, g$ and $\tilde g$ are new tau-functions,
and then impose the bilinear equations among these tau-functions
$$D_yf\cdot \tilde g -{1\over 2\kappa}(f\tilde g-\tilde fg)=0, \eqno(2.22a)$$
$$D_y\tilde f\cdot g -{1\over 2\kappa}(\tilde fg-f\tilde g)=0. \eqno(2.22b)$$
 Obviously, the tau-functions $F, \widetilde F, G$ and $\widetilde G$ specified in (2.21) satisfy  (2.14).
 If we substitute (2.21) into (2.15) and use (2.22), we can show that the bilinear equation (2.15) is
 satisfied automatically. 
Under these settings, the following proposition holds. \par
\bigskip
\noindent{\bf Proposition 2.1.} {\it Assume the relations (2.21) and (2.22). Then,
the bilinear equations (2.20) reduce to the bilinear equations for $f, \tilde f, g$ and $\tilde g$
$$D_\tau D_yf\cdot \tilde g -{1\over 2\kappa}D_\tau f\cdot \tilde g +{1\over 2\kappa}D_\tau \tilde f\cdot g-\kappa (f\tilde g -\tilde fg)=0, \eqno(2.23a)$$
$$D_\tau D_y\tilde f\cdot g -{1\over 2\kappa}D_\tau \tilde f\cdot g +{1\over 2\kappa}D_\tau f\cdot \tilde g-\kappa (\tilde fg -f\tilde g)=0. \eqno(2.23b)$$}
\par
\noindent {\bf Proof.}  We first use (2.21) to derive the following identities which are verified easily by direct computation:
$$D_\tau \widetilde F\cdot F=f\tilde g D_\tau \tilde f\cdot g-\tilde fgD_\tau f\cdot \tilde g,$$
$$D_\tau D_y G\cdot G=2f\tilde g D_\tau D_yf\cdot \tilde g -2(D_\tau f\cdot \tilde g)(D_yf\cdot \tilde g).$$
If we substitute these into (2.20$a$), use (2.22$a$) to eliminate a term $D_yf\cdot \tilde g$ and then divide the resultant expression by $2G$, we arrive at
the bilinear equation (2.23$a$). 
The derivation of (2.23$b$) can be done in the same way.  \hspace{\fill} $\square$ \par
\bigskip
The system of bilinear equations (2.22) and (2.23) is more tractable than the original system (2.15) and (2.20) since the former system
has no constraint on the tau-functions such as (2.14). \par
\bigskip
\noindent{\it 2.3. Parametric representation for the $N$-soliton solution}\par
\bigskip
\noindent Here, we provide an explicit parametric representation for the $N$-soliton solution of the mCH equation. 
The following two theorems refer to the main results.
\par
\bigskip
\noindent {\bf Theorem 2.1.} {\it The mCH equation (1.1) admits the parametric representation
$$u(y, \tau)={1\over 2 {\rm i}\kappa}\left({\rm ln}\,{\tilde f\tilde g\over fg}\right)_\tau, \eqno(2.24a)$$
$$x(y, \tau)={y\over\kappa}+{\rm ln}\,{\tilde g g\over \tilde ff}+d, \eqno(2.24b)$$
where the tau-functions $f, \tilde f, g$ and $\tilde g$ solve the system of bilinear equations (2.22) and (2.23) and $d$ is an
arbitrary constant.}\par
\bigskip
\noindent {\bf Proof.}  The expression (2.24{\it a}) is a consequence of  (2.12{\it a}) and (2.21). To derive (2.24{\it b}),
we deduce from  (2.12{\it b}),  (2.21) and  (2.22) that
$$\cos\,\phi=1+\kappa\left({\rm ln}\,{\tilde gg\over \tilde ff}\right)_y.$$
We substitute this relation into (2.11) and perform the integral with respect to $y$ 
under the boundary conditions $f, \tilde f, g, \tilde g \rightarrow 1$ as $y\rightarrow -\infty$
which are consistent with the boundary conditions for $F, \widetilde F, G$ and $\widetilde G$. 
Then, the expression (2.24{\it b}) follows immediately.   The constancy of $d$ has already been demonstrated. \hspace{\fill} $\square$ \par
\bigskip
\noindent {\bf Theorem 2.2.} {\it  The tau-functions $f, \tilde f, g$ and $\tilde g$ constituting the $N$-soliton solution are given respectively by the
expressions
 $$f=\sum_{\mu=0,1}{\rm exp}\left[\sum_{j=1}^N\mu_j\left(\xi_j+\psi_j+{\pi\over 2}\,{\rm i}\right)
+\sum_{1\le j<k\le N}\mu_j\mu_k\gamma_{jk}\right], \eqno(2.25a)$$
$$\tilde f=\sum_{\mu=0,1}{\rm exp}\left[\sum_{j=1}^N\mu_j\left(\xi_j+\psi_j-{\pi\over 2}\,{\rm i}\right)
+\sum_{1\le j<k\le N}\mu_j\mu_k\gamma_{jk}\right], \eqno(2.25b)$$
  $$g=\sum_{\mu=0,1}{\rm exp}\left[\sum_{j=1}^N\mu_j\left(\xi_j-\psi_j+{\pi\over 2}\,{\rm i}\right)
+\sum_{1\le j<k\le N}\mu_j\mu_k\gamma_{jk}\right], \eqno(2.25c)$$
 $$\tilde g=\sum_{\mu=0,1}{\rm exp}\left[\sum_{j=1}^N\mu_j\left(\xi_j-\psi_j-{\pi\over 2}\,{\rm i}\right)
+\sum_{1\le j<k\le N}\mu_j\mu_k\gamma_{jk}\right], \eqno(2.25d)$$
where
$$\xi_j=k_j\left(y-{2\kappa^3\over 1-(\kappa k_j)^2}\,\tau-y_{j0}\right), \qquad (j=1, 2, ..., N),\eqno(2.25e)$$
$${\rm e}^{\gamma_{jl}}=\left({k_j-k_l\over k_j+k_l}\right)^2, \qquad (j, l=1, 2, ..., N; j\not=l),
\eqno(2.25f)$$
$$e^{-\psi_j}=\sqrt{1-\kappa k_j\over 1+\kappa k_j},\qquad (j=1, 2, ..., N).\eqno(2.25g)$$
Here, $k_j$ and $y_{j0}$ are arbitrary complex parameters satisfying the conditions $k_j\not= k_l$
for $j\not=l$, ${\rm Re}\,k_j>0$ for all $j$, and $N$ is an arbitrary positive integer. The notation $\sum_{\mu=0,1}$
implies the summation over all possible combinations of $\mu_1=0, 1, \mu_2=0, 1, ..., 
\mu_N=0, 1$}.\par 
\bigskip
In the appendix, we show by means of mathematical induction
 that $f, \tilde f, g$ and $\tilde g$ from (2.25) solve the system of bilinear equations (2.22) and (2.23).
The $N$-soliton solution given by (2.24) with (2.25) is characterized by the $2N$ complex parameters $k_j$ and $y_{j0}\ (j=1, 2. ..., N)$. 
The parameters $k_j$ determine the amplitude and the velocity of the  solitons, whereas
the parameters $y_{j0}$ determine the position (or phase) of the solitons. We have imposed the conditions ${\rm Re}\,k_j>0\ (j=1, 2, ..., N)$ to satisfy the
boundary conditions for the tau-functions $f, \tilde f, g$ and $\tilde g$.  
In the following analysis, we consider the real solutions which are realized simply by imposing the conditions $\tilde f=f^*$ and $\tilde g=g^*$, where the asterisk
 denotes complex conjugate.
 \par
 The parametric solution (2.24) would become a multi-valued (or singular) function of $x$, as was the case for the short-pulse [15] and generalized sG equations [18, 19],
 unless we impose certain conditions on the parameters $k_j ( j=1, 2, .., N)$.  To establish a criterion for obtaining single-valued (or smooth) functions, we require that the mapping
 (2.2) is one-to-one which demands $x_y>0$. It follows from (2.9{\it a}), (2.12{\it b}) and (2.21) with $\tilde f=f^*, \tilde g=g^*$ that this condition yields the inequality
 $$|{\rm Re}\,fg^*|> |{\rm Im}\,fg^*|. \eqno(2.26)$$
Although it is difficult in general to extract the condition for the parameters $k_j$ from (2.26) for the  $N$-soliton solution, we will give it explicitly in the case of the one-soliton solution. \par
\bigskip
\noindent{\bf Remark 2.1.} The parametric representation for the $N$-soliton solution of the mCH equation has the  structure similar to that of the generalized sG equation 
$$u_{tx}=(1-\partial_x^2)\,\sin\,u,\qquad u=u(x, t). \eqno(2.27)$$
Actually, it can be written in the form [18]
$$u(y, \tau)={\rm i}\,{\rm ln}\,{\tilde f\tilde g\over fg}, \eqno(2.28a)$$
$$x(y, \tau)=y+\tau+{\rm ln}\,{\tilde gg\over \tilde ff}+y_0. \eqno(2.28b)$$
Here, the tau-functions $f, \tilde f, g$ and $\tilde g$  follow from (2.25)  with the identification  $\kappa=1, k_j=p_j$, and  by 
replacing the $\tau$ dependence as $\tau/p_j^2$ for $j=1, 2, ..., N$ so that $\xi_j=p_j\{y+(1/p_j^2)\tau-y_{j0}\}$, where $p_j$ are complex parameters.
See expressions (2.30), (2.35) and (2.36) as well as remark 2.5 in [18]. The implication of this interesting observation  will be considered in a separate context. \par
\bigskip
\noindent{\bf Remark 2.2.}\ By using an elementary theory of determinants, we can show that the tau-functions $f, \tilde f, g$ and $\tilde g$ from (2.25) solve the
system of bilinear equations (2.22) and (2.23). To this end, we first 
shift the phase variables $\xi_j$ as $\xi_j\rightarrow \xi_j-\psi_j\ (j=1, 2, ..., N)$ and then
express the $\tau$ and $y$ derivatives of $f$ and $\tilde f$ as well as those of $g$ and $\tilde g$ in terms of the bordered
determinants following the procedure developed for the $N$-soliton solution of the CH and generalized sG equations [18, 25].
Then, the  proof of (2.22) follows from the result given in the appendix of [18]. Now, we differentiate (2.22a)\, ((2.22b)) by $\tau$ and add the resultant expression to (2.23a)\,((2.23b)) to obtain
 the alternative bilinear equations
$$\left(f_{\tau y}-{1\over 2\kappa}\,f_\tau-\kappa^2f_y\right)\tilde g -(f_\tau-\kappa^2f)\tilde g_y +{1\over 2\kappa}\tilde f_\tau g=0, \eqno(2.29a)$$
$$\left(\tilde f_{\tau y}-{1\over 2\kappa}\,\tilde f_\tau-\kappa^2\tilde f_y\right)g -(\tilde f_\tau-\kappa^2\tilde f)g_y +{1\over 2\kappa}f_\tau \tilde g=0. \eqno(2.29b)$$
 With the aid of Jacobi's formula as well as a few basic formulas for determinants, we can verify that equations (2.29) are satisfied with the tau-functions (2.25).
The computation is performed straightforwardly but it is  too lengthy to reproduce  here. \par
\bigskip
\noindent{\bf Remark 2.3.}\ The tau-functions $f, \tilde f, g$ and $\tilde g$ from (2.25) have the same structure as that of the $N$-soliton solution of a model
equation for shallow-water waves [21]
$$q_\tau+2\kappa^3q_y+4\kappa^2qq_\tau-2\kappa^2q_y\int_y^\infty q_\tau dy-\kappa^2q_{\tau yy}=0,\qquad q=q(y, \tau). \eqno(2.30)$$
To see this, we shift the phase variables $\xi_j$ as $\xi_j\rightarrow \xi_j-\psi_j-\pi{\rm i}/2\ (j=1, 2. ..., N)$ in (2.25$a$), for example,
and then introduce the dependent variable transformation $q=-2({\rm ln}f)_{yy}$. It turns out that $q$ solves equation (2.30) [21, 28]. 
We point out that the tau-function $f$ thus obtained is the basic constituent for the $N$-soliton solutions of the CH [22-25] and  
dispersionless mCH [9] equations.  Thus, at the level of the tau-functions, the $N$-soliton solutions for these equations have a common structure.
\par
\bigskip
\noindent{\bf 3. Properties of soliton solutions}\par
\bigskip
\noindent In this section, we describe the properties of soliton solutions.  
We show that a variety of solutions arise from the parametric representation (2.24) with (2.25) in accordance with the values of the soliton parameters.
First, we deal with the one-soliton solutions which include the smooth soliton, symmetric
singular soliton and antisymmetric singular soliton. Subsequently, we address the two-soliton solutions such as the smooth two-soliton and  a smooth soliton
and  symmetric singular soliton pair, 
 as well as a breather  which stems from the smooth two-soliton solution as a degenerate case. 
Finally, we provide  the formula for the phase shift of the $N$-soliton solution. \par
\bigskip
\noindent{\it 3.1. One-soliton solution}\par
\medskip
\noindent{\it 3.1.1. Smooth soliton.}\
 The tau-functions $f$ and $g$ corresponding to the one-soliton solution are given by (2.25) with $N=1$. Explicitly,
$$f=1+{\rm i}\,{\rm e}^{\xi+\psi}, \eqno(3.1a)$$
$$g=1+{\rm i}\,{\rm e}^{\xi-\psi}, \eqno(3.1b)$$
with
$$\xi=k\left(y-{2\kappa^3\over 1-(\kappa k)^2}\,\tau-y_0\right), \eqno(3.1c)$$
$${\rm e}^{-\psi}=\sqrt{1-\kappa k\over 1+\kappa k}, \eqno(3.1d)$$
where we have put $\xi=\xi_1, \psi=\psi_1, k=k_1$ and $y_0=y_{10}$ for simplicity. 
The boundary conditions for $f$ and $g$, i.e., $f, g \rightarrow 1$ as $y\rightarrow -\infty$ require that the real part of $k$ is positive.
Since we are concerned with the real one-soliton solutions, we assume that all the parameters are real. The complex parameters will be introduced for constructing the breather
solutions. \par
The parametric representation of the one-soliton solution follows by introducing (3.1) into (2.24). We write it in the form
$$u={4\kappa^2 k\over  \{1-(\kappa k)^2\}^{3/2}}{\cosh\,\xi\over \cosh\,2\xi+{1+(\kappa k)^2\over 1-(\kappa k)^2}}, \eqno(3.2a)$$
$$X\equiv x-ct-x_0={\xi\over \kappa k}+{\rm ln}{1-\kappa k\,\tanh\,\xi\over 1+\kappa k\,\tanh\,\xi}, \eqno(3.2b)$$
with
$$c={2\kappa^2\over 1-(\kappa k)^2}, \eqno(3.2c)$$
where $c$ is the velocity of the soliton in the $(x, t)$ coordinate system,  $x_0=y_0/\kappa$
and the constant $d$ in (2.24b) has been chosen appropriately such that $\xi=0$ corresponds to $X=0$. \par
The $X$ derivative of $u$ can be computed by using the relation $u_X=u_\xi/X_\xi$, which gives
$$u_X=-{4\kappa^3k^2\over \{1-(\kappa k)^2\}^{3/2}}\,{\sinh\,\xi\over \cosh\,2\xi+{1+(\kappa k)^2\over 1- (\kappa k)^2}}. \eqno(3.3)$$
It can be checked by direct substitution that the parametric solution (3.2) indeed satisfies equation (1.1).
\par
The smooth soliton solution
is obtainable if one imposes a certain condition on the parameter $k$ which can be derived from (2.26) and (3.1).  Alternatively, we compute
the quantity $x_y$ directly from (3.2{\it b}) and obtain
$$x_y={1\over\kappa}\left[1-{4(\kappa k)^2\over 1-(\kappa k)^2}\,{1\over \cosh\,2\xi+{1+(\kappa k)^2\over 1-(\kappa k)^2}}\right]. \eqno(3.4)$$
The condition $x_y>0$ must hold for arbitrary value of $\xi$ to assure the smoothness of the solution.  This leads to the inequality
$$0<\kappa k<{1\over\sqrt{2}}. \eqno(3.5)$$
\par

\begin{figure}[t]
\begin{center}
\includegraphics[width=10cm]{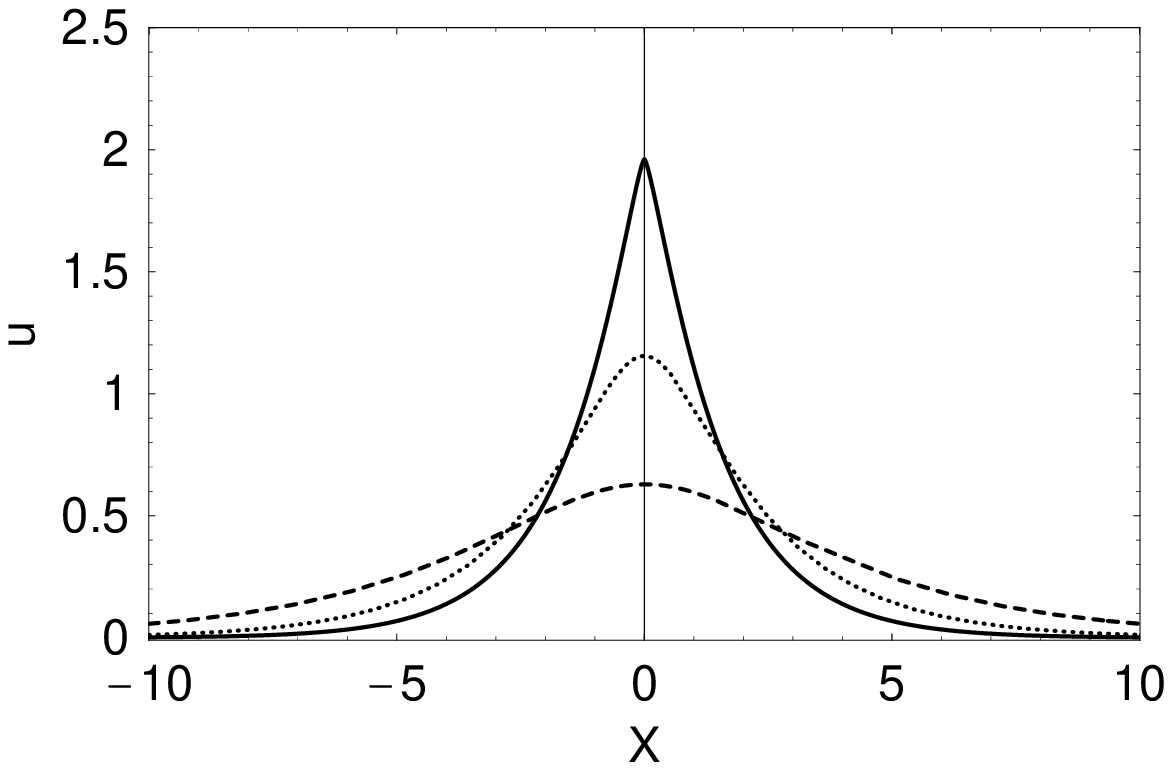}
\end{center}
\noindent {{\bf Figure 1.}\ The profile of  smooth solitons with $\kappa=1$: $\kappa k=0.3$\,(dashed curve), $\kappa k=0.5$\,(dotted curve), $\kappa k=0.7$\,(solid curve).}
\end{figure}

The smooth one-soliton solution represents a bright soliton whose center position $x_c$ locates at $x_c=ct+x_0$  and has the amplitude $A$ given by
$$A=\sqrt{2(c-2\kappa^2)}. \eqno(3.6)$$
This amplitude-velocity relation  follows immediately by eliminating the parameter $k$ from the amplitude $A=u|_{\xi=0}=2\kappa^2k/\{1-(\kappa k)^2\}^{1/2}$ and the velocity $c$ given by (3.2{\it c}).
The inequality (3.5) restricts allowable values of $c$ and $A$. To be more specific,
$2\kappa^2<c<4\kappa^2,\ 0<A<2\kappa$. \par
\begin{figure}[t]
\begin{center}
\includegraphics[width=10cm]{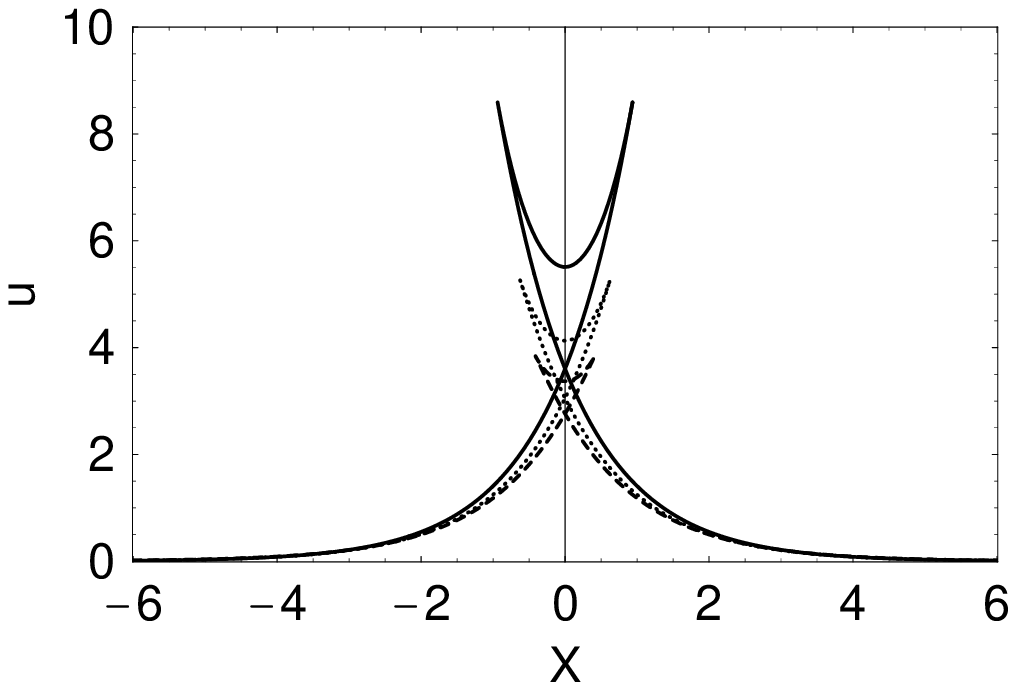}
\end{center}
\noindent {{\bf Figure 2.}\ The profile of  symmetric singular solitons with $\kappa=1$: $\kappa k=0.86$\,(dashed curve), $\kappa k=0.90$\,(dotted curve), $\kappa k=0.94$\,(solid curve).}
\end{figure}
Figure 1 depicts the profile of smooth solitons against the stationary variable $X$ defined by (3.2{\it b}) for three distinct values of $\kappa k$ with $\kappa=1$.
As the value of the parameter $\kappa k$ increases, the amplitude grows and the width narrows. When it tends to the upper limit $\kappa k=1/\sqrt{2}$ of the inequality (3.5), 
then the smoothness of the solution is lost at the crest of the soliton.  To see this in more detail, we expand $u$ and $X$ near the crest $\xi=0$.
Specifically, when $\kappa k=1/\sqrt{2}$, the leading terms of the expansions read
$$u=2\kappa\left[1-{1\over 8}\,\xi^4+O(\xi^6)\right], \eqno(3.7a)$$
$$X={1\over 3\sqrt{2}}\,\xi^3+O(\xi^5). \eqno(3.7b)$$
By eliminating the variable $\xi$ from (3.7), we find that the profile $u$ of the soliton  near the crest $X=0$
is approximated by
$$u=2\kappa \left[1-{(3\sqrt{2})^{4/3}\over 8}\,X^{4/3}+O(X^2)\right]. \eqno(3.8)$$
We can see from (3.8) that  the $n$th derivative of $u$ with respect to $X$ does not
exist for $n\geq 2$. 
This novel feature of the solution is striking contrast to the usual peakon which has a discontinuous first derivative at the crest. However, the solitonic nature
of the peaked solution presented here must be justified after its stability has been established.
\par
A similar structure to this solution has been observed in the analysis of the smooth soliton solution of the dispersionless
mCH equation where  a constant background field  plays the role of the parameter $\kappa$ [9]. \par
\bigskip
\noindent{\it 3.1.2. Symmetric singular soliton.}\
The singular solitons with a symmetric profile exist in the range of the parameter $1/\sqrt{2}<\kappa k<1$.
In Figure 2, the profile of symmetric singular solitons is depicted  for three distinct values of $\kappa k$ with $\kappa=1$.
 They exhibit two crests and become  three-valued functions of $X$ in the range $-X_0<X<0,\ 0<X<X_0$, where $X_0$ will be specified below. 
At the origin $X=0$, $u$  takes two values $u_1$ and $u_2$\ $(0<u_1<u_2)$. We can show that if $1/\sqrt{2}<\kappa k<1$, then 
the coordinate $X=X(\xi)$ from (3.2{\it b}) has three zeros $\xi=0, \pm \xi_1$ where the value of $\xi_1$ can be computed numerically.
 Consequently, $u_1=u|_{\xi=\pm \xi_1}$ and $u_2=A$ with $A$ being given by (3.6). 
 The maximum value $u_{max}$ of the amplitude is attained 
 at $X=\pm X_0\ (\xi=\mp\xi_0)$, where $\xi_0=\tanh^{-1}[\sqrt{2(\kappa k)^2-1}/\kappa k]$ and
 $$X_0=-{1\over 2\kappa k}\,{\rm ln}\,{\kappa k+\sqrt{2(\kappa k)^2-1}\over \kappa k-\sqrt{2(\kappa k)^2-1}}+{\rm ln}\,{1+\sqrt{2(\kappa k)^2-1}\over 1-\sqrt{2(\kappa k)^2-1}}. $$
 Then, $u_{max}= c/2\kappa$. 
 The slope $u_X$ at $X=\pm X_0$ is evaluated simply by putting $\xi=\mp\xi_0$ in (3.3), which gives
 $\pm\sqrt{(2(\kappa k)^2-1)/2(1-(\kappa k)^2)}$. \par
 Finally, it is instructive to take the small dispersion limit $\kappa\rightarrow 0$ with $c$ being fixed. 
 This limiting procedure is called the peakon limit. It has been used successfully to produce the peakons from the smooth solitons of the CH [30, 31] and Degasperis-Procesi (DP) [32] equations.
 In view of (3.2{\it c}), we must take the limit
 $\kappa k \rightarrow 1$ simultaneously. It turns out that $u_{max}\rightarrow \infty,\ u_X\rightarrow \pm\infty\ (X_0\rightarrow \pm\infty)$, showing that  the two crests located at $X=\pm X_0$ 
 tend to $\pm\infty$ and their profile evolves into  cusp. Note that, in this limit,  $u_1\rightarrow \sqrt{c/2}$ and $u_2\rightarrow \sqrt{2c}$. 
 We recall that a similar singular solution has been obtained  for the dispersionless mCH equation [9]. 
 Thus, unlike the W-shaped singular soliton of the Novikov equation  which reduces to a peakon 
 in an appropriate limit (see figure 4 of [20]), the symmetric singular soliton under consideration does not recover the peakon
  obtained in  [6]. 
 \par
\bigskip
\noindent{\it 3.1.3. Antisymmetric singular soliton.}\
A novel type of singular soliton appears in  the parameter range $\kappa k>1$.  Such singular solitons can be
constructed from the smooth solitons if one shifts the phase variable $\xi$ as $\xi\rightarrow \xi+\pi{\rm i}/2$, or equivalently
replaces the phase constants $x_0$ and $y_0$ by $x_0-\pi {\rm i}/(2\kappa k)$ and $y_0-\pi {\rm i}/(2k)$, respectively.
In this setting, $\cosh\,\xi\rightarrow {\rm i}\,\sinh\,\xi, \cosh\,2\xi\rightarrow -\cosh\,2\xi$ and $\tanh\,\xi \rightarrow \coth\,\xi$,
giving rise to  the parametric representation of the singular soliton solution
$$u={4\kappa^2 k\over \{(\kappa k)^2-1\}^{3/2}}\,{\sinh\,\xi\over \cosh\,2\xi+{(\kappa k)^2+1\over (\kappa k)^2-1}}, \eqno(3.9a)$$
$$X\equiv x-ct-x_0={\xi\over \kappa k}+{\rm ln}\,{\kappa k-\tanh\,\xi\over \kappa k+\tanh\,\xi}. \eqno(3.9b)$$
The expression (3.9) becomes an antisymmetric function of $X$ as evidenced from the relations  $u(-\xi)=-u(\xi)$ and $X(-\xi)=-X(\xi)$.
It follows from (3.9) that
$$u_X=-{4\kappa^3k^2\over \{(\kappa k)^2-1\}^{3/2}}\,{\cosh\,\xi\over \cosh\,2\xi+{(\kappa k)^2+1\over (\kappa k)^2-1}}. \eqno(3.10)$$
Thus, $u_X$  takes only   negative and finite values, showing that contrary to the symmetric singular soliton, the profile always exhibits  negative slope.
 \par
\begin{figure}[t]
\begin{center}
\includegraphics[width=10cm]{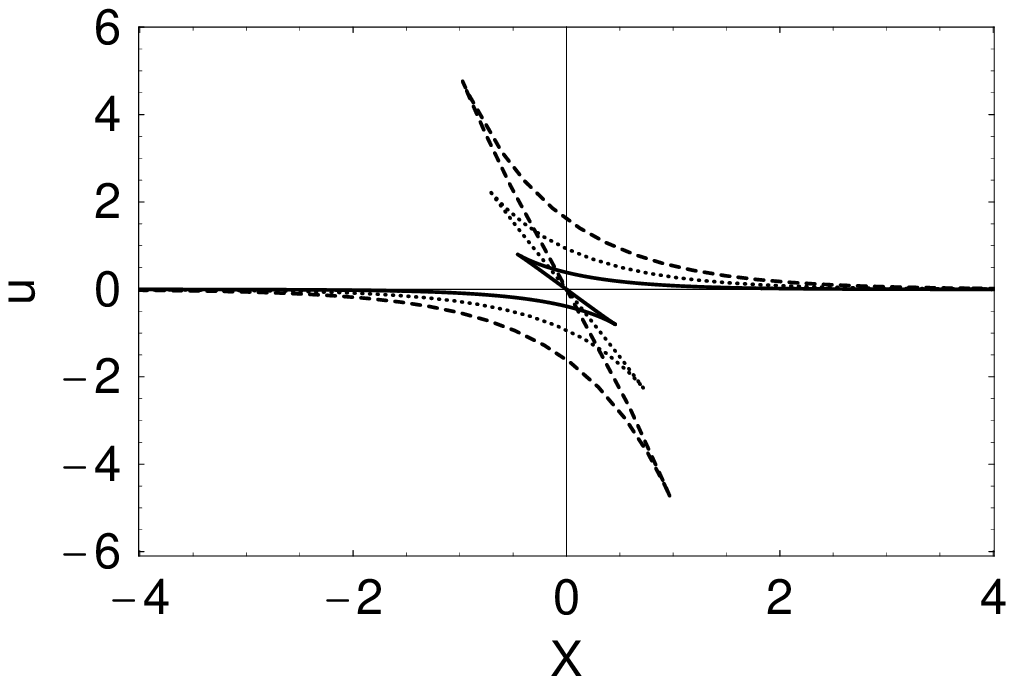}
\end{center}
\noindent {{\bf Figure 3.}\ The profile of  antisymmetric singular solitons with $\kappa=1$: $\kappa k=1.1$\,(dashed curve), $\kappa k=1.2$\,(dotted curve), $\kappa k=1.5$\,(solid curve).}
\end{figure}
 Figure 3 depicts the profile of antisymmetric singular solitons  for three distinct values of $\kappa k$ with $\kappa=1$. We can see from (3.9{\it a}) that $u$ attains the maximum (minimum) value
$\kappa/\{(\kappa k)^2-1\}$ (-$\kappa/\{(\kappa k)^2-1\}$) at $\xi=-\xi_2$ ($\xi=\xi_2$), where $\xi_2=\tanh^{-1}(\kappa k/\sqrt{2(\kappa k)^2-1})$. 
The value of $X$ corresponding to $\xi_2$, which is denoted by $X_2$, is found from (3.9{\it b}) as
$$X_2=-{1\over 2\kappa k}\,{\rm ln}\,{\sqrt{2(\kappa k)^2-1}+\kappa k\over \sqrt{2(\kappa k)^2-1}-\kappa k}+{\rm ln}\,{\sqrt{2(\kappa k)^2-1}+1\over \sqrt{2(\kappa k)^2-1}-1}. \eqno(3.11)$$
In the interval $-X_2<X<X_2$, $u$ becomes a three-valued function of $X$. 
In the limit of $\kappa k\rightarrow 1$, the positions $\pm X_2$ of the two crests move to infinity and their amplitudes grow indefinitely. \par
We recall that the parametric solution (3.9) has been obtained recently in classifying the traveling-wave solutions of the mCH equation [29].
However, the detailed analysis of the solution is presented here for the first time. \par
\bigskip
\noindent{\it 3.2. Two-soliton solution}\par
\medskip
\noindent The tau-functions (2.25) with $N=2$ for the two-soliton solutions can be written  in the form
$$ f=1+{\rm i}({\rm e}^{\xi_1+\psi_1}+{\rm e}^{\xi_2+\psi_2})-\left(k_1-k_2\over k_1+k_2\right)^2{\rm e}^{\xi_1+\xi_2+\psi_1+\psi_2}, \eqno(3.12a)$$
$$ g=1+{\rm i}({\rm e}^{\xi_1-\psi_1}+{\rm e}^{\xi_2-\psi_2})-\left(k_1-k_2\over k_1+k_2\right)^2{\rm e}^{\xi_1+\xi_2-\psi_1-\psi_2}. \eqno(3.12b)$$
\par
The parametric solution (2.24) with (3.12) exhibits a variety of  solutions describing the interaction of two solitons. Here, we consider the 
three types of solutions which are composed  of two smooth solitons, one smooth soliton and one  symmetric singular soliton and a breather, respectively.
The other types of solutions will be dealt with elsewhere.   
\par
\medskip
\noindent{\it 3.2.1. Smooth soliton - smooth soliton.} The smooth two-soliton solution is obtained if one chooses the real parameters $k_j$ subjected to
the conditions $0<\kappa k_j<1/\sqrt{2}, (j=1, 2)$. As already pointed out in remark 2.1, the structure of the tau-functions (3.12) is
the same as that of the two-soliton solutions of the generalized sG equation except for the $\tau$ dependence. 
The asymptotic  analysis of the solution mimics that of the generalized sG equation.
Hence, we omit it and outline the  results (see section 3.2 of [18] for details). \par

\begin{figure}[t]
\begin{center}
\includegraphics[width=15cm]{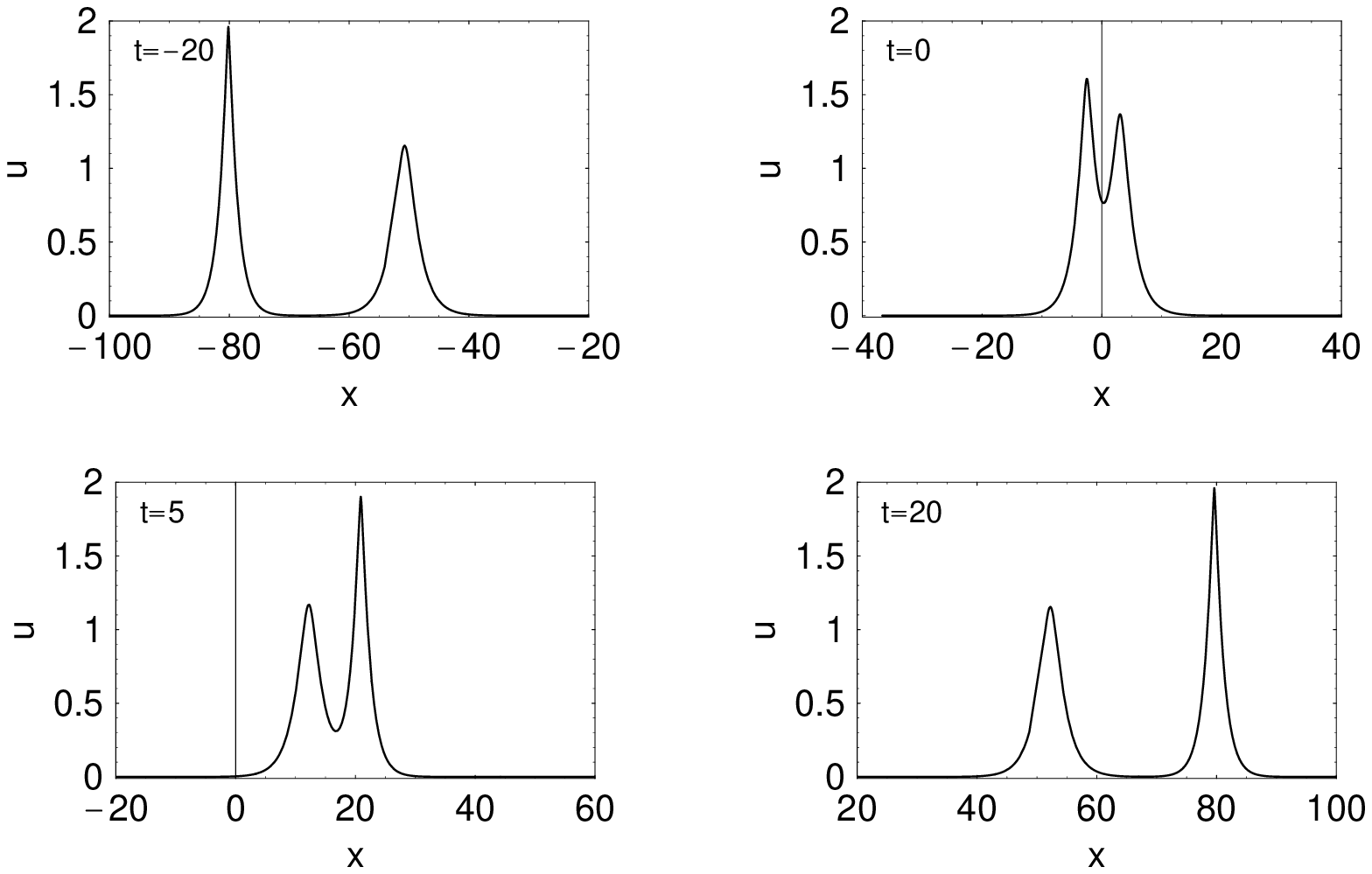}
\end{center}
\noindent {{\bf Figure 4.}\ The interaction between two smooth solitons with the parameters $\kappa=1, \kappa k_1=0.7, \kappa k_2=0.5$ and $y_{10}=y_{20}=0$.}
\end{figure}

 Figure 4 illustrates the interaction  of two smooth solitons for four distinct values of $t$.
This figure shows clearly the solitonic nature of the solution.  The asymptotic state of the solution for large time 
is represented by a superposition of two smooth solitons, each of which has the profile given by (3.2). The net effect of 
the interaction is the phase shift. Let $\Delta_1$ and $\Delta_2$ be the phase shifts of the large and small solitons, respectively. 
It then follows from the expressions
 (3.6) of [18] with $p_1=\kappa k_1$ and $p_2=\kappa k_2\ (0<\kappa k_2<\kappa k_1<1/\sqrt{2})$ that
 $$\Delta_1=-{1\over \kappa k_1}\ln\left({k_1-k_2\over k_1+k_2}\right)^2-\ln\left({1+\kappa k_2\over 1-\kappa k_2}\right)^2, \eqno(3.13a)$$
 $$\Delta_2={1\over \kappa k_2}\ln\left({k_1-k_2\over k_1+k_2}\right)^2+\ln\left({1+\kappa k_1\over 1-\kappa k_1}\right)^2. \eqno(3.13b)$$
  In the illustrated example, the amplitude of the large soliton is 1.96 whereas that of the small soliton is 1.16. The phase shifts are
  evaluated by the formulas (3.13), giving $\Delta_1=2.92  $ and $\Delta_2=-3.70$. \par
    The first terms of (3.13) coincide with the corresponding formulas for the KdV and sG equations whereas the second terms originate from
  the coordinate transformation (2.2). Recall that the above formulas for the phase shifts are exactly the same as those for the two-soliton
  solution of the CH equation (see, for example, [25]). 
  We first summarise the features of the phase shift of the CH two-soliton solution and then proceed to the mCH case. \par
  In the CH case, the allowable values of the parameters are restricted by the inequality
  $0<\kappa k_2<\kappa k_1<1$. 
    The detailed analysis reveals that the large  soliton is always shifted forwards
  ($\Delta_1>0)$ whereas  the sign of $\Delta_2$ depends on the values of $\kappa k_1$ and $\kappa k_2$.
       There arise three cases for the allowable values of the phase shifts: (i) $\Delta_2<0 \leq \Delta_1$, (ii) $0<\Delta_2\leq \Delta_1$, (iii) $0<\Delta_1\leq  \Delta_2$.
     We can show that if $\kappa k_1<\kappa k_{1c}$, then $\Delta_2$ always takes a negative value (case (i) above), where 
    $\kappa k_{1c}=0.8336$ is a root of the transcendental equation $-4/(\kappa k_1)+{\rm ln}[(1+\kappa k_1)^2/(1-\kappa k_1)^2]=0$
which follows from (3.13{\it b}) by taking the limit $\kappa k_2\rightarrow 0$.
The cases (ii) and (iii) exhibit quite
 peculiar characteristics which have never been observed in the interaction process of the KdV and sG solitons. 
  \begin{figure}[t]
\begin{center}
\includegraphics[width=6cm]{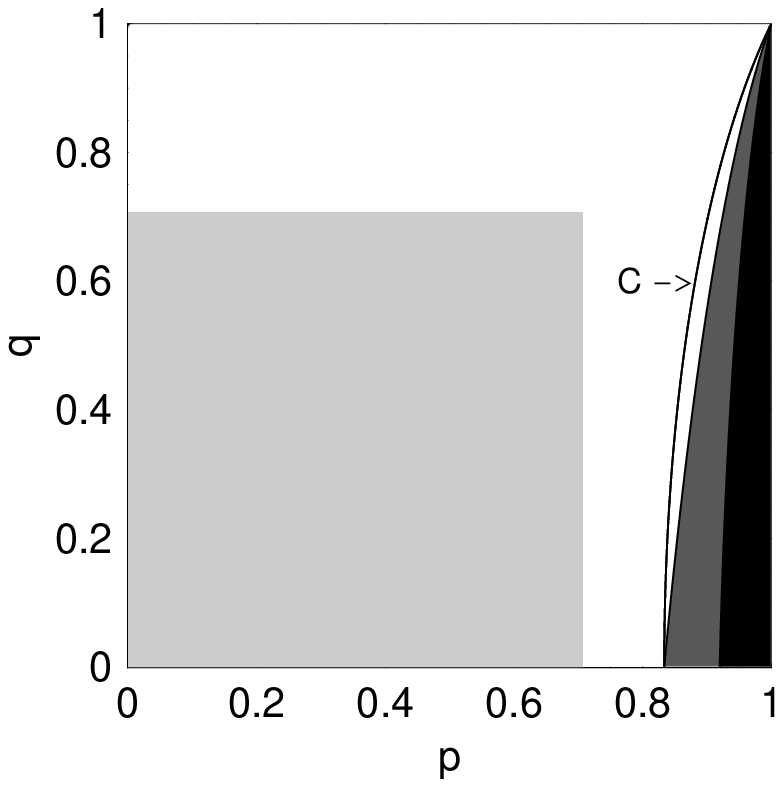}
\end{center}
\noindent {{\bf Figure 5.}\ The phase shift diagram in $(p, q)$ plane with $p=\kappa k_1$ and $q=\kappa k_2$. The critical curve $C$
separates the two regions  $\Delta_2<0$ (left)  and $\Delta_2>0$ (right).} The allowable values of $\kappa k_1$ and $\kappa k_2$ 
for the smooth solitons lie in the light gray region. 
\end{figure}

  \begin{figure}[t]
\begin{center}
\includegraphics[width=15cm]{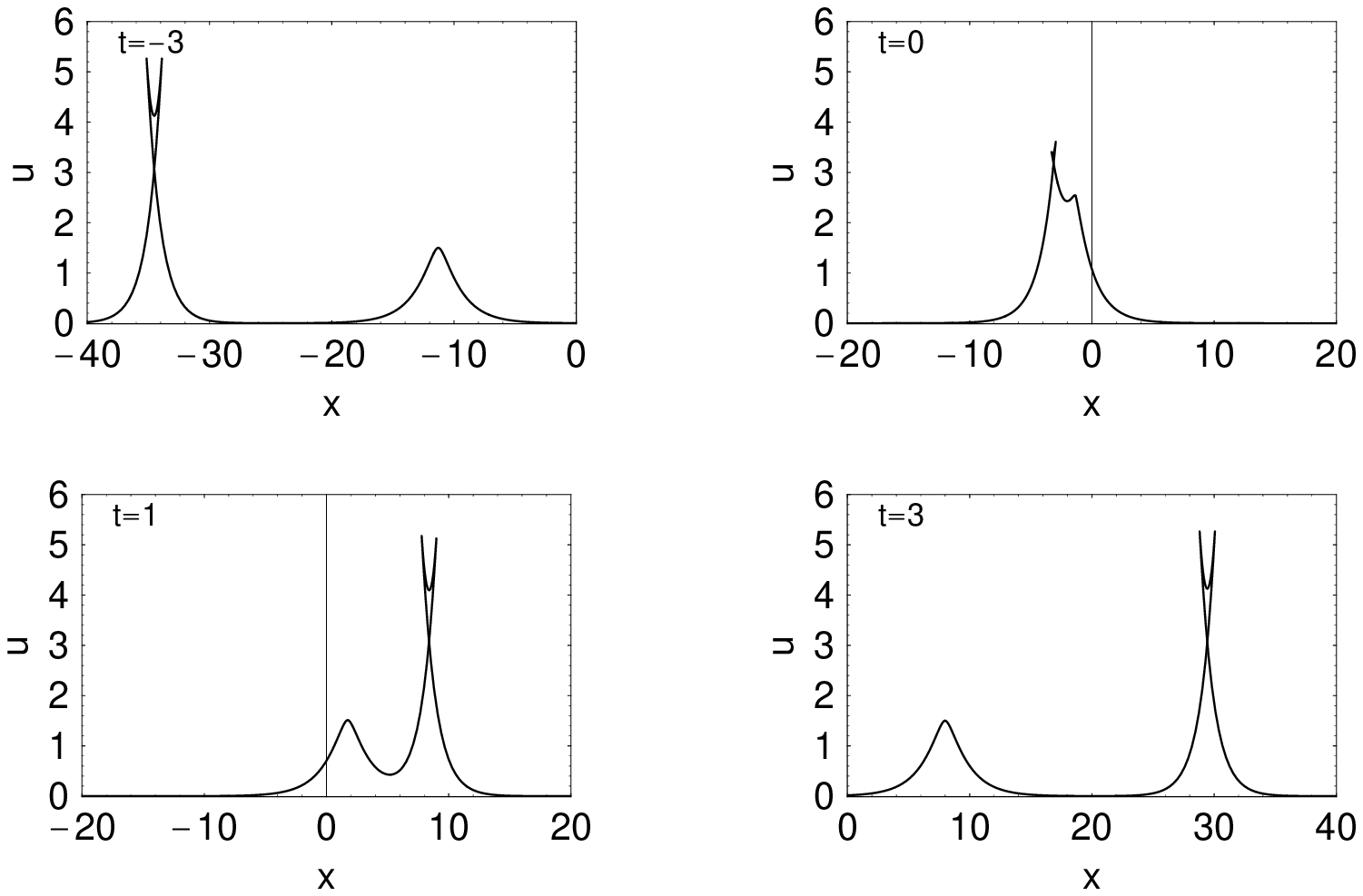}
\end{center}
\noindent {{\bf Figure 6.}\ The interaction between a smooth soliton ($k_2=0.6, y_{20}=0$) and a symmetric singular soliton ($k_1=0.9, y_{10}=0$).
The parameter $\kappa$ is set to 1 for both solitons.}
\end{figure}
Figure 5 plots the critical curves  in the $(p, q)$ plane with $p=\kappa k_1$ and $q=\kappa k_2$ which separate the above
three cases.  Note that the allowable values of $p$ and $q$ are restricted by the inequality $0<q<p<1$. 
For any pair $(p, q)$ lying in the  left region of the curve C,  the phase shifts satisfy the inequality indicated in case (i).
Notice that the curve C starts from the point $(\kappa k_{1c}, 0)$, increases monotonically and ends at the point $(1, 1)$. The phase shift $\Delta_2$ for the  small soliton
becomes zero along this curve.
The narrow region surrounded by the curve C and the curve separating the white and gray regions corresponds to case (ii)  whereas 
both the   gray and black regions correspond to case (iii). In particular, on the boundary separating the gray and black regions, the relation
$\Delta_2-\Delta_1=1$ holds. See also figure 2 of [32] which depicts an analogous diagram for the phase shift of the two-soliton solution of the DP  equation.
\par
    Now, in the mCH case considered here, since $\kappa k_1<1/\sqrt{2}<0.8336$, we see from figure 5 that $\Delta_2<0$, implying that
the small soliton is always shifted backward after the interaction of solitons. Thus, the behavior of smooth solitons is similar to that of the KdV and sG solitons despite
 the different structure of the formulas for the phase shifts. \par
\bigskip
    \noindent{\it 3.2.2. Smooth soliton - symmetric singular soliton.}  The two-soliton solution composed of a smooth soliton and a symmetric singular soliton 
    is obtained if one sets the parameters so that the inequalities
$1/\sqrt{2}<\kappa k_1<1$ and  $0<\kappa k_2<1/\sqrt{2}$ are satisfied. Figure 6 illustrates the interaction process  for four distinct values of $t$.
The solitonic feature of the solution is apparent from the figure. Actually, we can see that the singular soliton overtakes and emerges ahead of  the smooth soliton.
 After the interaction,  both solitons appear
without changing their profiles and suffer only the phase shifts, which can be evaluated by making use of (3.13).  
The characteristic of the interaction process differs from that of the smooth two-soliton case. 
Indeed, if $\kappa k_1<\kappa k_{1c}$, then $\Delta_2<0$ whereas if $\kappa k_{1c}<\kappa k_1$, then the allowable value of $\Delta_2$ is classified into
either  case (ii) or case (iii) mentioned above in accordance the value of $\kappa k_2$.
In the present example, $\Delta_1=0.804$ and $\Delta_2=0.524$ (case (ii)). \par
\bigskip
\noindent{\it 3.2.3. Breather.} The breather solution has a localized structure which oscillates with time and decays exponentially in space.
In the sG model, it can be interpreted as the bound state of a kink and an antikink. We show that in the mCH equation, the corresponding breather solution is
produced from the two-soliton solution by specifying the complex conjugate pair for the parameters. \par
Now, we put
$$k_1=a+{\rm i}b,\qquad k_2=a-{\rm i}b\ (=k_1^*), \qquad a>0, \eqno(3.14a)$$
$$y_{10}=\eta+{\rm i}\delta, \qquad y_{20}=\eta-{\rm i}\delta\ (=y_{10}^*). \eqno(3.14b)$$
Then, the tau-functions $f$ and $g$ from (3.12) reduce to
$$ f=1+{\rm i}({\rm e}^{\xi_1+\psi_1}+{\rm e}^{\xi_1^*+\psi_1^*})+\left({b\over a}\right)^2{\rm e}^{\xi_1+\xi_1^*+\psi_1+\psi_1^*}, \eqno(3.15a)$$
$$ g=1+{\rm i}({\rm e}^{\xi_1-\psi_1}+{\rm e}^{\xi_1^*-\psi_1^*})+\left({b\over a}\right)^2 {\rm e}^{\xi_1+\xi_1^*-\psi_1-\psi_1^*}, \eqno(3.15b)$$
 where
 $$\xi_1=\theta+{\rm i}\chi, \eqno(3.16a)$$
$$\theta=a(y-\nu_1\tau)-a\eta+b\delta,\qquad \nu_1={2\kappa^3\{1-\kappa^2(a^2+b^2)\}\over \{1-\kappa^2(a^2-b^2)\}^2+4\kappa^4(ab)^2}, \eqno(3.16b)$$
$$\chi=b(y-\nu_2\tau)-b\eta-a\delta,\qquad \nu_2={2\kappa^3\{1+\kappa^2(a^2+b^2)\}\over \{1-\kappa^2(a^2-b^2)\}^2+4\kappa^4(ab)^2}, \eqno(3.16c)$$
$${\rm e}^{-\psi_1}=\sqrt{1-\kappa^2(a^2-b^2)+2{\rm i}\kappa^2ab\over (1+\kappa a)^2+(\kappa b)^2}\equiv \alpha{\rm e}^{-{\rm i}\beta}. \eqno(3.16d)$$
 In terms of the new variables defined by (3.16), the tau-functions $f$ and $g$ can be rewritten as
$$f=1+2{\rm i}\alpha^{-1}{\rm e}^{\,\theta} \cos(\chi+\beta)+\alpha^{-2}\left({b\over a}\right)^2{\rm e}^{2\theta}, \eqno(3.17a)$$
$$g=1+2{\rm i}\alpha\,{\rm e}^{\,\theta} \cos(\chi-\beta)+\alpha^2\left({b\over a}\right)^2{\rm e}^{2\theta}. \eqno(3.17b)$$
\par
The smooth breather solutions are produced if we choose the parameters $a$ and $b$ such that condition (2.26) is satisfied.
As in the case of the corresponding problem for the generalized sG equation, it is not easy to find the allowable values of $a$ and $b$ analytically. However, an
inspection reveals that if the ratio $a/|b|$ is sufficiently small compared to 1, then the regularity of the solution would be assured. Figure 7 depicts the time evolution of a
smooth breather solution for  four distinct values of $t$. The breather propagates to the right while changing its profile and whose characteristic is similar to that
of the breather solution of the generalized sG equation [18]. \par
\bigskip

  \begin{figure}[t]
\begin{center}
\includegraphics[width=15cm]{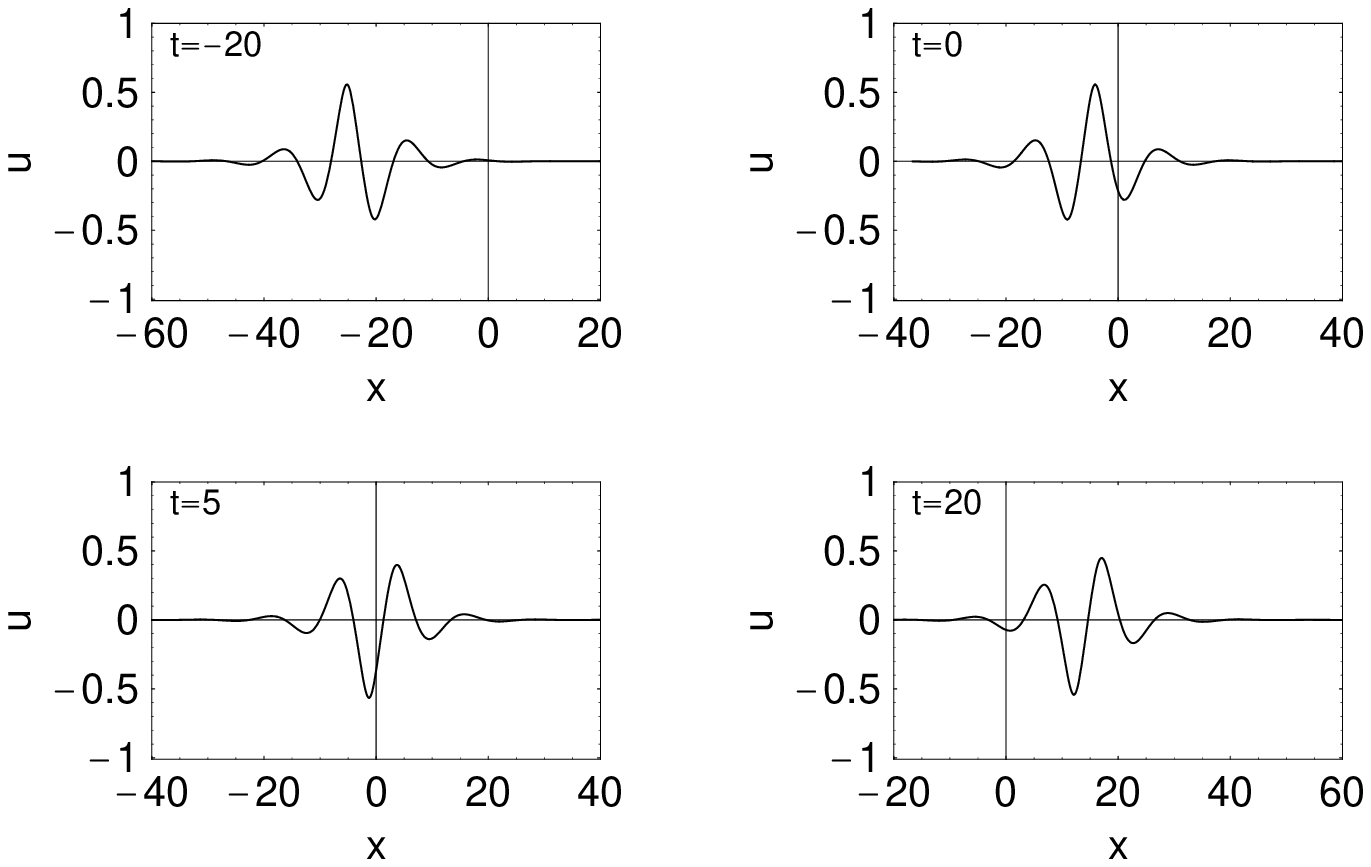}
\end{center}
\noindent {{\bf Figure 7.} The time evolution of a breather solution with the parameters $\kappa=1, a=0.2, b=0.5$ and $\eta=\delta=0$.}
\end{figure}

\noindent{\it 3.3. $N$-soliton solution}\par
\medskip
\noindent
The general multisoliton solutions are classified in accordance 
 values of the parameters $\kappa k_j\ (j=1, 2, ..., N)$.
Actually, the constituents of the solutions are composed of the smooth solitons, symmetric singular solitons and antisymmetric singular solitons   as well as
the  breathers.
The asymptotic analysis of the general $N$-soliton solution will not be performed here since  the similar analysis has been done
for the $N$-soliton solution of the generalized sG equation [18]. Here, we provide the formula for the phase shift. To this end, let us order the magnitude of
the parameters $\kappa k_j$ as
$0<\kappa k_N<\kappa k_{N-1}< ... <\kappa k_1$ so that the velocity of each soliton satisfies the condition $0<c_N<c_{N-1}< ... <c_1$ by (3.2c), where
$c_j=2\kappa^2/\{1-(\kappa k_j)^2\}\,(j=1, 2, ..., N)$.
 Then, the phase shift of the $n$th soliton is given by the formula
$$\Delta_n={1\over \kappa k_n}\left\{\sum_{j=1}^{n-1}\ln\left({k_n-k_j\over k_n+k_j}\right)^2
-\sum_{j=n+1}^N\ln\left({k_n-k_j\over k_n+k_j}\right)^2\right\}$$
$$+\sum_{j=1}^{n-1}\ln\left({1+\kappa k_j\over 1-\kappa k_j}\right)^2-\sum_{j=n+1}^{N}\ln\left({1+\kappa k_j\over 1-\kappa k_j}\right)^2,
\quad (n = 1, 2, ..., N). \eqno(3.18)$$
 See formula (3.26) of [18] with the identification $p_j=\kappa k_j\ (j=1, 2, ..., N)$. 
 If we restrict the largest parameter $\kappa k_1$ as $\kappa k_1<1/\sqrt{2}$, then the above formula  gives the phase shift for  $N$ interacting smooth solitons.
If $\kappa k_{N-m+1}<1/\sqrt{2}$ and $1/\sqrt{2}<\kappa k_{N-m}<\kappa k_{N-m-1}< ... <\kappa k_1<1$ , for example, then the asymptotic state of the solution for large time
is composed of $m$ smooth solitons and $N-m$ symmetric singular solitons.  In this specific case, formula (3.18) gives the phase shift of the smooth solitons for $1\leq n\leq m$
and that of the singular symmetric solitons for $m+1\leq n\leq N$, respectively.
It is also possible to construct the pure multibreather solutions as well as the  multisoliton-multibreather solutions 
following the recipe described in section 3.2.3. See section 3.3 of [18] for details. \par
\bigskip
\noindent {\bf 4. Reduction to the short pulse equation}\par
\bigskip
\noindent The SP equation (1.2) was obtained for the first time in an attempt to construct integrable differential equations associated with
pseudospherical surfaces [33].
Later, it was proposed as an alternative model to the cubic nonlinear Schr\"odinger (NLS) equation [14]. 
In the context of self-focusing of ultra-short pulses in nonlinear media, its validity would be beyond the scope of applicability of
the NLS equation which has been derived on the assumption of a slowly varying envelope approximation. See the recent review articles [17, 34]
 for the short pulse equation and related topics. Here, we demonstrate that the SP  equation, its $N$-soliton solution and formula for the
phase shift are all recovered from  the mCH equation under an appropriate scaling limit, or  the short-wave limit.
Note that the similar limiting procedure has been performed in [18, 19] for the generalized sG equation, leading to  the same results. \par
\bigskip
\noindent {\it 4.1.  Short-wave limit of the mCH equation} \par
\bigskip
\noindent The mCH equation is reducible to the SP equation in the short-wave limit [6]. Here, we
demonstrate it for completeness. 
We recall that the similar limiting procedure has been undertaken for the CH and DP equations [35]. \par
First, we introduce the scaling variables 
 in accordance with the relations
$$u=\epsilon^2 \bar u, \quad x=\epsilon \bar x, \quad  y=\epsilon \bar y,\quad   t={\bar t \over \epsilon}, \quad  \tau ={\bar\tau\over\epsilon},$$
$$ k_j={\bar k_j\over\epsilon}, \quad y_{j0}=\epsilon \bar y_{j0}\ (j=1, 2, ...,N), \quad d=\epsilon \bar d. \eqno(4.1)$$
where $\epsilon$ is  a small parameter. 
Rewriting the derivatives in terms of the new variables, the mCH equation (1.1) is recast to
$$\epsilon(\epsilon^2\bar u-\bar u_{\bar x\bar x})_{\bar t}+2\kappa^2\epsilon\bar u_{\bar x}
+{1\over \epsilon}[(\epsilon^2\bar u-\bar u_{\bar x\bar x})( \epsilon^4\bar u^2-\epsilon^2\bar u_{\bar x}^2)]_{\bar x}=0. \eqno(4.2)$$
If we expand $\bar u$ in powers of $\epsilon$ as $\bar u=\bar u_0+\epsilon \bar u_1 + ...$ and substitute it into (4.2),  we obtain, at the 
 the leading order of the expansion, the equation for $\bar u_0$ 
$$-\bar u_{0,\bar t\bar x\bar x}+2\kappa^2\epsilon\bar u_{0,\bar x}+[\bar u_{0,\bar x\bar x}\bar u_{0,\bar x}^2]_{\bar x}=0. \eqno(4.3)$$
If we put $\bar v=\bar u_{0,\bar x}$ in (4.3), we arrive, after dropping the bar attached to the variables $t, x$ and $v$, at
the SP equation (1.2). \par
\bigskip
\noindent{\it 4.2. Short-wave limit of the $N$-soliton solution}\par
\medskip
\noindent First, shift the phase variables $\xi_j$ as  $\xi_j\rightarrow \xi_j-\psi_j\ (j=1, 2, ..., N)$ and then take the limit  $\epsilon \rightarrow 0$. The tau-functions $f$  and $\tilde f$
from (2.25$a$) and (2.25$b$), respectively have  
the limiting forms, which are given by
 $$f\rightarrow \bar f= \sum_{\mu=0,1}{\rm exp}\left[\sum_{j=1}^N\mu_j\left(\bar\xi_j+{\pi\over 2}\,{\rm i}\right)
+\sum_{1\le j<k\le N}\mu_j\mu_k\bar\gamma_{jk}\right], \eqno(4.4a)$$
$$\tilde f\rightarrow \bar{\tilde f}= \sum_{\mu=0,1}{\rm exp}\left[\sum_{j=1}^N\mu_j\left(\bar\xi_j-{\pi\over 2}\,{\rm i}\right)
+\sum_{1\le j<k\le N}\mu_j\mu_k\bar\gamma_{jk}\right], \eqno(4.4b)$$
   with
   $$\bar\xi_j=\bar k_j\left(\bar y+{2\kappa \over\bar k_j^2}\,\bar\tau -\bar y_{j0}\right), \quad (j=1, 2, ..., N),\eqno(4.4c)$$
   $${\rm e}^{\bar\gamma_{jl}}=\left({\bar k_j-\bar k_l\over \bar k_j+\bar k_l}\right)^2, \qquad (j, l=1, 2, ..., N; j\not=l). \eqno(4.4d)$$
   To perform the limiting procedure for the tau-function $g$, we need to retain terms up to order $\epsilon$.
   Using  the expansion
   \begin{align*}
   {\rm exp}\left(-2\sum_{j=1}^N\mu_j\psi_j\right)
   &= \prod_{j=1}^N\left({1-\kappa k_j\over 1+\kappa k_j}\right)^{\mu_j}  \\
  &= {\rm exp}\left(-\pi {\rm i}\sum_{j=1}^N\mu_j\right)\left(1-2\epsilon \sum_{j=1}^N{\mu_j\over \kappa \bar k_j}\right)+O(\epsilon^2), \tag{4.5}
  \end{align*}
    the tau-function $g$ from (2.25$c$) can be developed in powers of $\epsilon$ as
    \begin{align*}
    g &= \sum_{\mu=0, 1}\left(1-2\epsilon \sum_{j=1}^N{\mu_j\over \kappa \bar k_j}\right)
   {\rm exp}\left[\sum_{j=1}^N\mu_j\left(\bar\xi_j-{\pi\over 2}{\rm i}\right)
+\sum_{1\le j<k\le N}\mu_j\mu_k\bar\gamma_{jk}\right]+O(\epsilon^2) \\
&=\bar{\tilde f}-{\epsilon\over \kappa^2}\bar{\tilde f}_{\bar\tau}+O(\epsilon^2). \tag{4.6a}
\end{align*}
The corresponding expansion of the tau-function $\tilde g$ from (2.25$d$) reads
$$\tilde g=\bar f-{\epsilon\over \kappa^2}\bar f_{\bar\tau}+O(\epsilon^2). \eqno(4.6b)$$
\par
Now, the relation (2.10) has the leading order expansion
$$\bar v\equiv \bar u_{0,\bar x}
=-\bar\phi_{\bar\tau}/(2\kappa), \eqno(4.7)$$ 
where we have put $\phi=\bar\phi$. The scaling variable $\bar\phi$ has a limiting form $\bar\phi=2{\rm i}\,{\rm ln}(\bar{\tilde f}/\bar f)$ by virtue of 
(2.12$b$), (2.21), (4.4) and (4.6) which, substituted in (4.7), gives the expression of $\bar v$ in terms of the tau-functions $\bar f$ and $\bar{\tilde f}$:
$$\bar v=-{\rm i\over\kappa}\left({\rm ln}\,{\bar{\tilde f}\over \bar f}\right)_{\bar \tau}. \eqno(4.8)$$
\par
Similarly, it follows from (4.4) and  (4.6) that
 \begin{align*}
{\rm ln}\,{\tilde gg\over \tilde ff}
&={\rm ln}\left[{\left(\bar f-{\epsilon\over \kappa^2}\bar f_{\bar\tau}+O(\epsilon^2)\right) 
\left(\bar{\tilde f}-{\epsilon\over \kappa^2}\bar{\tilde f}_{\bar\tau}+O(\epsilon^2)\right)\over \bar{\tilde f}\bar f}\right] \\
&= -{\epsilon\over \kappa^2}({\rm ln}\,\bar{\tilde f}\bar f)_{\bar \tau} +O(\epsilon^2). \tag{4.9}
\end{align*}
Introducing the scaling variables $\bar x, \bar y$ and $\bar d$ from (4.1) as well as (4.9) into (2.24b), we obtain the limiting form of $x$:
$$\bar x={\bar y\over \kappa}-{1\over\kappa^2}\,({\rm ln}\,\Bar{\Tilde f}\bar f)_{\bar \tau}+\bar d. \eqno(4.10)$$
The expressions (4.8) and (4.10) coincide with the parametric representation for the $N$-soliton solution of the SP equation [15]. \par
\bigskip
\noindent {\bf Remark 4.1.} Performing the short-wave limit to the bilinear equations (2.22) and (2.23) with use of (4.4) and (4.6), 
they reduce to the following system of bilinear equations for
$\bar f$ and $\bar{\tilde f}$:
$$D_{\bar\tau}D_{\bar y}\bar f\cdot \bar f=\kappa(\bar f^2-\bar{\tilde f}^2), \eqno(4.11a)$$
$$D_{\bar\tau}D_{\bar y}\bar{\tilde f}\cdot \Bar{\tilde f}=\kappa(\bar{\tilde f}^2-\bar f^2). \eqno(4.11b)$$
Recall that the system of equations (4.11) is a bilinear form of the sG equation.
Actually, the sG equation $\bar u_{\bar\tau\bar y}=\,{\rm sin}\,\bar u$ can be 
transformed to the bilinear equations (4.11) through the dependent variable
transformation $\bar u=2{\rm i}\,{\rm ln}(\Bar{\Tilde f}/\bar f)$. \par
\bigskip
\noindent{\it 4.3. Short-wave limit of the phase shift}\par
\medskip
\noindent The short-wave limit of  formula (3.18) for the phase shift can be performed simply.
Indeed, the scaling $\Delta_n=\epsilon\bar\Delta_n$ of the phase shift and that of the parameters $k_j$ given by (4.1) lead,
after taking the limit $\epsilon\rightarrow 0$,  to the 
phase shift of the $n$th soliton 
$$\bar\Delta_n={1\over \kappa \bar k_n}\left\{\sum_{j=1}^{n-1}\ln\left({\bar k_n-\bar k_j\over \bar k_n+\bar k_j}\right)^2
-\sum_{j=n+1}^N\ln\left({\bar k_n-\bar k_j\over \bar k_n+\bar k_j}\right)^2\right\}$$
$$+\sum_{j=1}^{n-1}{4\over\kappa \bar k_j}-\sum_{j=n+1}^{N}{4\over\kappa \bar k_j},
\quad (n = 1, 2, ..., N). \eqno(4.12)$$
This formula recovers formula for the phase shift of the $N$-soliton solution of the SP equation presented in [15]. \par
\bigskip
\noindent{\bf Remark 4.2.} Under the scaling transformations
$$u=\epsilon \bar u, \quad x-t=\epsilon \bar x, \quad  y=\epsilon \bar y,\quad   t={\bar t \over \epsilon}, \quad  \tau ={\bar\tau\over\epsilon},$$
$$ k_j={\bar k_j\over\epsilon}, \quad y_{j0}=\epsilon \bar y_{j0}\ (j=1, 2, ...,N), \quad d=\epsilon \bar d, \eqno(4.13)$$
the generalized sG equation (2.27) reduces to the SP equation  (1.2) in the limit of $\epsilon\rightarrow 0
$. Indeed, rewriting (2.27) in terms of the new scaling variables introduced in (4.13) and using the Taylor series expansion of the function $\sin\,\epsilon\bar u$, 
we can develop (2.27) to
$$\epsilon\left(\bar u_{\bar t\bar x}-{1\over\epsilon^2}\bar u_{\bar x\bar x}\right)=\epsilon\bar u-{\epsilon^3\over 6}\,\bar u^3+ \cdots
 -{1\over\epsilon^2}\left(\epsilon\bar u-{\epsilon^3\over 6}\,\bar u^3+{\epsilon^5\over 120}\,\bar u^5\right)_{\bar x\bar x}.$$
 The leading terms of order $\epsilon$ in the above expansion yield the SP equation (1.2).
 Note that the terms of order $\epsilon^{-1}$ are canceled each other.
The $N$-soliton solution
(2.28)  recovers the parametric representation  (4.8) and  (4.10 ) with  (4.4) 
for the $N$-soliton solution of the SP equation   whereas  formula  (3.18) for the phase shift  
reduces to  formula (4.12). \par
\bigskip
\leftline{\bf 5. Concluding remarks} \par
\bigskip
\noindent In this paper, a systemtic method has been developed for solving the mCH equation under the rapidly decreasing boundary condition.
We have obtained a variety of  solutions which include the smooth and singular solitons and breathers, and investigated their
properties.  
The existence of the smooth soliton
solutions of the mCH equation was found to be restricted to a certain range of the amplitude parameter $\kappa k$. The same situation has been encountered
in the analysis of the smooth soliton solutions of the dispersionless mCH equation subjected to the nonvanishing boundary condition [9]. To be more specific, for the former case, the 
allowable range of the parameter is  $0<\kappa k<1/\sqrt{2}$ (see (3.5)) whereas for the latter case, this is given by $0<u_0 k<\sqrt{3}/2$, where $u_0$ is a constant background field
such that $u\rightarrow u_0$ as $|x|\rightarrow 0$. Thus, the peakon limit, i.e.,  $\kappa k\rightarrow 1
(u_0 k\rightarrow 1)$ with the velocity of the soliton being fixed, is not relevant for these smooth solitons.
This is in striking contrast to the peakon limit of the smooth solitons of the CH and DP equations [30-32] as well as that of the Novikov equation [20],
for which the limiting procedure  recovered peakons.
On the other hand, the peakon solution of the form $u=\sqrt{3c/2}\,{\rm exp}(-|x-ct-x_0|)$ has been shown to exist for 
the dispersionless mCH equation (1.1) with $\kappa=0$ [6].
A possible way to recover the peakon is to take the peakon limit for the singular soliton. Indeed, this procedure has been applied to the symmetric singular soliton
of the Novikov equation [20].
Unfortunately,  a similar procedure has not succeeded for the current problem, as already shown here  for the 
symmetric singular soliton (section 3.1.2).
We  will postpone this interesting issue  for a future study. \par
The exact method of solution developed here will be applied to construct soliton solutions of a variant of the mCH equation 
$$ m_t+2\kappa^2u_x+\alpha_1[m(u^2-u_x^2)]_x+\alpha_2(2mu_x+m_xu)=0,\quad  m=u-u_{xx},\quad u=u(x, t), \eqno(5.1)$$
where $\alpha_1$ and $\alpha_2$ are arbitrary constants. 
This equation is a linear combination of the CH and mCH equations. 
The particular cases $\alpha_1=0$ and $\alpha_2=0$ reduce to the CH and mCH equations, respectively.
It is an integrable generalization of the Gardner equation which is a linear combination of the KdV and
modified KdV equations [1-3]. Actually, the integrability of equation (5.1) was established recently by constructing the Lax pair [36]. 
While the smooth and singular single soliton solutions of traveling-wave type were obtained in [36] for equation (5.1),
the  multisoliton solutions are not  yet available. The various problems mentioned above will be dealt with
in subsequent papers. \par
\bigskip
\noindent{\bf Acknowledgements} \par
\bigskip
\noindent The author would like to thank the referees for  useful suggestions and comments. \par
\bigskip
\leftline{\bf Appendix. Proof of (2.22) and (2.23)} \par
\bigskip
\noindent
In this appendix, we show that the tau-functions (2.25)
 solve the bilinear equations (2.22) and (2.23). We use a mathematical induction  similar to that used for the $N$-soliton solution of the DP and Novikov equations [20, 37].
 We first prove (2.22$a$) and then proceed to (2.23{\it a}). 
 The proof of (2.22{\it b}) and (2.23{\it b}) can be done in the same way and hence it will be omitted.  \par
 \bigskip
 \noindent {\it A.1. Proof of (2.22a)} \par
 \bigskip
 \noindent Substituting the tau-functions $f, \tilde f, g$ and $\tilde g$ from (2.25) into the bilinear equation (2.22{\it a}) and using
 the formula
$$D_\tau^mD_y^n\,{\rm exp}\left[\sum_{i=1}^N\mu_i\xi_i\right]\cdot {\rm exp}\left[\sum_{i=1}^N\nu_i\xi_i\right]$$
$$=\left\{-\sum_{i=1}^N(\mu_i-\nu_i)k_i\tilde c_i\right\}^m\left\{\sum_{i=1}^N(\mu_i-\nu_i)k_i\right\}^n{\rm exp}\left[\sum_{i=1}^N(\mu_i+\nu_i)\xi_i\right],\qquad (m, n=0, 1, 2, ...),$$
 where $\tilde c_i=2\kappa^3/\{1-(\kappa k_i)^2\}$, 
 the identity to be proved becomes
 $$\sum_{\mu, \nu=0,1}\left[\left\{\sum_{i=1}^N(\mu_i-\nu_i)k_i-{1\over 2\kappa}\right\}\,{\rm exp}\left[{\pi\over 2}{\rm i}\sum_{i=1}^N(\mu_i-\nu_i)\right]
 +{1\over 2\kappa}\,{\rm exp}\left[-{\pi\over 2}{\rm i}\sum_{i=1}^N(\mu_i-\nu_i)\right]\right]$$
 $$\times {\rm exp}\left[\sum_{i=1}^N(\mu_i+\nu_i)\xi_i+\sum_{i=1}^N(\mu_i-\nu_i)\psi_i+\sum_{1\leq i<j\leq N}(\mu_i\mu_j+\nu_i\nu_j)\gamma_{ij}\right]=0. \eqno(A.1)$$
\par
Let $P_{m,n}$ be the coefficient of the factor ${\rm exp}\left[\sum_{i=1}^n\xi_i+\sum_{i=n+1}^m2\xi_i\right]\ (1\leq n<m\leq N)$ on the left-hand side of (A.1). Correspondingly, the
summation with respect to $\mu_i$ and $\nu_i$ must be performed under the conditions
$$\mu_i+\nu_i=1\qquad (i=1, 2, ..., n), \qquad \mu_i=\nu_i=1\qquad (i=n+1, n+2, ..., m),$$
$$\mu_i=\nu_i=0\qquad (i=m+1, m+2, ..., N). \eqno(A.2)$$
To proceed, it is crucial to introduce the new summation indices $\sigma_i$ by the relations $\mu_i=(1+\sigma_i)/2,\ \nu_i=(1-\sigma_i)/2$ for $i=1, 2, ..., n$, where $\sigma_i$ takes either the value $+1$ or $-1$. 
It turns out that $\mu_i\mu_j+\nu_i\nu_j=(1+\sigma_i\sigma_j)/2$. \par
Now, under conditions (A.2), we deduce that
$$\sum_{1\leq i<j\leq N}(\mu_i\mu_j+\nu_i\nu_j)\gamma_{ij}={1\over 2}\sum_{1\leq i<j\leq n}(1+\sigma_i\sigma_j)\gamma_{ij}+\sum_{i=1}^m\sum_{\substack{j=n+1\\ (j\not=i)}}^m\gamma_{ij}. \eqno(A.3)$$
Using (A.3), $P_{m,n}$ can be written in the form
$$P_{m,n}=\sum_{\sigma=\pm 1}\left[\left\{\sum_{i=1}^n\sigma_ik_i-{1\over 2\kappa}\right\}{\rm exp}\left[{\pi\over 2}{\rm i}\sum_{i=1}^n\sigma_i\right]
+{1\over 2\kappa}{\rm exp}\left[-{\pi\over 2}{\rm i}\sum_{i=1}^n\sigma_i\right]\right]$$
$$\times {\rm exp}\left[\sum_{i=1}^n\sigma_i\psi_i+{1\over 2}\sum_{1\leq i<j\leq n}(1+\sigma_i\sigma_j)\gamma_{ij}+\sum_{i=1}^m\sum_{\substack{j=n+1\\ (j\not=i)}}^m\gamma_{ij}\right]. \eqno(A.4)$$
The following relations stem from (2.25{\it f}), (2.25{\it g}) and the definition of $\sigma_i$:
$${\rm exp}\left[{1\over 2}\sum_{1\leq i<j\leq n}(1+\sigma_i\sigma_j)\gamma_{ij}\right]
=\prod_{1\leq i<j\leq n}{(\sigma_ik_i-\sigma_jk_j)^2\over (k_i+k_j)^2}, $$
$${\rm exp}\left[\sum_{i=1}^n\sigma_i\psi_i\right]
=\prod_{i=1}^n{1+\kappa \sigma_i k_i\over (1-(\kappa k_i)^2)^{1/2}}, \qquad {\rm exp}\left[{\pi\over 2}{\rm i}\sum_{i=1}^n\sigma_i\right]={\rm i}^n\prod_{i=1}^n\sigma_i. \eqno(A.5)$$
If we insert (A.5) into (A.4) and drop a multiplicative factor independent of the summation indices $\sigma_i$, the identity to be proved reduces to
$$P_n(k_1, k_2, ..., k_n)\equiv \sum_{\sigma=\pm 1}\left[\sum_{i=1}^n\sigma_i k_i-{1\over 2\kappa}+{(-1)^n\over 2\kappa}  \right]\prod_{i=1}^n\sigma_i\prod_{i=1}^n(1+\kappa \sigma_ik_i)$$
$$\times \prod_{1\leq i<j \leq n} (\sigma_ik_i-\sigma_jk_j)^2=0, \qquad (n=1, 2, ..., N). \eqno(A.6)$$
Obviously, $P_n$ is a symmetric polynomial of $k_i\ (i=1, 2, ..., n)$ by virtue of the summation indices $\sigma_i\ (i=1, 2, ..., n)$ and is odd with respect to each $k_i$ due to the factor $\prod_{i=1}^n\sigma_i$.\par
The proof proceeds by mathermatical induction. The identity (A.6) can be proved easily for $n=1, 2$. Assume that $P_{n-2}=0$. 
First, we note that the relation
$$P_n|_{k_1=0}=\sum_{\sigma_1=\pm 1}\sigma_1\times({\rm terms\ independent\ of}\ \sigma_1)=0, \eqno(A.7)$$
holds because of the summation with respect to $\sigma_1$, showing that $k_1=0$ is a single zero of $P_n$.
Then,
$$P_n|_{k_1=k_2}=-8k_1^2(1-(\kappa k_1)^2)\prod_{i=3}^n(k_1^2-k_i^2)^2P_{n-2}(k_3, k_4, ..., k_n)=0, \eqno(A.8)$$
by the assumption of induction. On the other hand,
$${\partial P_n\over\partial k_1}=\sum_{\sigma=\pm 1}\prod_{i=2}^n\sigma_i\prod_{i=1}^n(1+\kappa \sigma_i k_i) \prod_{1\leq i<j \leq n} (\sigma_ik_i-\sigma_jk_j)^2$$
$$+\sum_{\sigma=\pm 1}\kappa\left[\sum_{i=1}^n\sigma_i k_i-{1\over 2\kappa}+{(-1)^n\over 2\kappa}  \right]\prod_{i=2}^n\sigma_i\prod_{i=2}^n(1+\kappa \sigma_i k_i) \prod_{1\leq i<j \leq n} (\sigma_ik_i-\sigma_jk_j)^2$$
$$+\sum_{\sigma=\pm 1}\left[\sum_{i=1}^n\sigma_i k_i-{1\over 2\kappa}+{(-1)^n\over 2\kappa}  \right]\prod_{i=1}^n\sigma_i\prod_{i=1}^n(1+\kappa \sigma_i k_i) 
{\partial\over\partial k_1}\left[\prod_{1\leq i<j \leq n} (\sigma_ik_i-\sigma_jk_j)^2\right]$$
$$=P_{n1}+P_{n2}+P_{n3}. \eqno(A.9)$$
We evaluate $P_{n1}$ at $k_1=k_2$ to obtain
$$P_{n1}\big|_{k_1=k_2}=-4k_1^2(1-(\kappa k_1)^2)\prod_{i=3}^n(k_1^2-k_i^2)^2\sum_{\sigma_1=\pm 1}\sigma_1 
 {\sum_{\sigma=\pm 1}}^{\prime\prime} \left[\prod_{i=3}^n\sigma_i(1+\kappa\sigma_i k_i)\prod_{3\leq i<j \leq n} (\sigma_ik_i-\sigma_jk_j)^2\right],\eqno(A.10)$$
where the notation ${\sum}^{\prime\prime}_{\sigma=\pm 1}$ implies the exclusion of $\sigma_1$ and $\sigma_2$ from the set of the indices $\{\sigma_1, \sigma_2, ..., \sigma_n\}$.
 Performing  the  summation with respect to $\sigma_1$, we eventually 
arrive at $P_{n1}\big|_{k_1=k_2}=0$.
Similarly, 
$$P_{n2}\big|_{k_1=k_2}=-4\kappa k_1^2\prod_{i=3}^n(k_1^2-k_i^2)^2\sum_{\sigma_1=\pm 1}\sigma_1(1-\kappa\sigma_1)P_{n-2}(k_3, k_4, ..., k_n)=0. \eqno(A.11)$$
Last, by a straightforward computation, we find that
$$P_{n3}\big|_{k_1=k_2}$$
$$=-4k_1(1-(\kappa k_1)^2)\left[\prod_{i=3}^n(k_1^2-k_i^2)^2+2k_1^2\sum_{i=3}^n(k_1^2-k_i^2)\prod_{\substack{j=3\\ (j\not=i)}}^n(k_1^2-k_j^2)^2\right]P_{n-2}(k_3, k_4, ..., k_n)$$
$$-8k_1^2(1-(\kappa k_1)^2)\sum_{\sigma_1=\pm 1}\sigma_1 {\sum_{\sigma=\pm 1}}^{\prime\prime}
\left[\sum_{i=3}^n\sigma_i k_i-{1\over 2\kappa}+{(-1)^n\over 2\kappa}  \right]$$
$$\times \prod_{i=3}^n\sigma_i\prod_{i=3}^n(1+\kappa \sigma_i k_i) 
\sum_{i=3}^n(k_1^2-k_i^2)\sigma_ik_i\prod_{\substack{j=3\\ (j\not=i)}}^n(k_1^2-k_j^2)^2\prod_{3\leq i<j \leq n} (\sigma_ik_i-\sigma_jk_j)^2. \eqno(A.12)$$
The first term on the right-hand side of (A.12) vanishes by the assumption of induction whereas the second term becomes zero due to the summation with respect to $\sigma_1$
and hence $P_{n3}\big|_{k_1=k_2}=0$. 
Thus, $\partial P_n/\partial k_1=0$ at $k_1=k_2$.  
By the similar argument, we confirm that the  equalities  $P_n=\partial P_n/\partial k_1=0$ hold at $k_1=-k_2$ as well. It turns out that $k_1=\pm k_2$ 
are double zeros of $P_n$. When coupled with (A.7), we see that  $P_n$ has a factor $k_1(k_1-k_2)^2(k_1+k_2)^2$.
Taking into account  the symmetry of $P_n$ in $k_i\,(i=1, 2, ..., n)$, the above result
reveals that $P_n$ can be factored by a polynomial
$$\prod_{i=1}^nk_i\prod_{1\leq i<j\leq n}(k_i^2-k_j^2)^2,$$
of  $k_i\,(i=1, 2, ..., n)$ of degree $2n^2-n$. On the other hand, the degree of $P_n$ from (A.6) is $n^2+1$ at most, which is impossible for $n\geq 2$ except  $P_n\equiv 0$.
This completes the proof of (2.22a).  \hspace{\fill} $\square$ \par
\bigskip
 \noindent {\it A.2. Proof of (2.23a)} \par
 \bigskip
\noindent The proof of (2.23{\it a})  parallels  that for (2.22{\it a}). Hence,  we omit the detail and outline the result. The expression corresponding to (A.6), which is denoted by $Q_n$, 
takes the form
$$Q_n(k_1, k_2, ..., k_n)\equiv \sum_{\sigma=\pm 1}\Biggl[\kappa\left\{-2\kappa \sum_{i=1}^n\sigma_i k_i+1-(-1)^n\right\}\sum_{i=1}^n\sigma_i k_i\prod_{\substack{j=1\\ (j\not=i)}}^n(1-(\kappa k_j)^2)$$
$$-\left\{1-(-1)^n\right\}\prod_{i=1}^n(1-(\kappa k_i)^2)\Biggr]
 \prod_{i=1}^n\sigma_i\prod_{i=1}^n(1+\kappa \sigma_ik_i) \prod_{1\leq i<j \leq n} (\sigma_ik_i-\sigma_jk_j)^2=0, \ (n=1, 2, ..., N). \eqno(A.13)$$
 \par
Now, the identitiy (A.13) holds for $n=1,2, 3$, as checked easily by direct computation.  Assume that $Q_{n-2}=0$. Then, we can show  that 
$$Q_n|_{k_1=0}=0, \qquad Q_n|_{k_1=\pm k_2}=0, \qquad {\partial Q_n\over \partial k_1}\big|_{k_1=\pm k_2}=0. \eqno(A.14)$$ 
The symmetry of $Q_n$ in $k_i\ (i=1, 2, ..., n)$ as well as (A.14) ensures that $Q_n$ has a factor 
$$\prod_{i=1}^nk_i\prod_{1\leq i<j\leq n}(k_i^2-k_j^2)^2,$$
whose degree in $k_i (i=1, 2, ..., n)$  is $2n^2-n$. On the other hand, the degree of $Q_n$ from (A.12) is $n^2+2n$ at most. This is impossible for $n\geq 4$ except $Q_n=0$.
Since (A.13) holds up to $n=3$, we conclude that $Q_n=0$ for all $n$, completing the proof of (2.23{\it a}). \hspace{\fill} $\square$ \par
\bigskip\
\newpage
\leftline{\bf Reference} \par
\baselineskip=5.5mm
\begin{enumerate}[{[1]}]
\item  Fokas A S 1995 On a class of physically important integrable equations  {\it Physica D} {\bf 87} 145-50
\item  Fuchssteiner B  1996 Some tricks from the symmetry-toolbox for nonlinear equations: Generalizations of the Camassa-Holm equation  {\it Phys. D} {\bf 95} 229-43 
\item  Olver P J and  Rosenau P 1996 Tri-Hamiltonian duality between solitons and solitary-wave solutions having compact support  {\it Phy. Rev. E} {\bf 53} 1900-6
\item  Camassa R and Holm D D 1993 An integrable shallow water equation with peaked solitons {\it Phys. Rev. Lett.} {\bf 71} 1661-4
\item  Qiao Z 2006 A new integrable equation with cuspons and W/M-shape-peaks solitons  {\it J. Math. Phys}. {\bf 47} 112701 
\item  Gui G,  Liu Y,  Olver P J and  Qu C 2013  Wave-breaking and peakons for a modified Camassa-Holm equation {\it  Commun. Math. Phys.} {\bf 319} 731-59
\item  Qiao Z and Li X Q 2011 An integrable equation with nonsmooth solitons {\it Theor. Math. Phys.} {\bf 167} 584-9
\item  Ivanov R I and  Lyons T 2012  Dark solitons of the Qiao's hierarchy  {\it J. Math. Phys.} {\bf 53} 123701
\item  Matsuno Y 2013 B\"acklund transformation and smooth multisoliton solutions for a modified Camassa-Holm equation with cubic nonlinearity {\it J. Math. Phys.} {\bf 54} 051504
\item  Bies P M, G\'orka P and Reyes E G 2012 The dual modified Korteweg-de Vries-Fokas-Qiao equation: Geometry and local analysis {\it J. Math. Phys}. {\bf 53} 073710
\item  Qu C, Liu X and Liu Y 2013 Stability of peakons for an integrable modified Camassa-Holm equation with cubic nonlinearity {\it  Commun. Math. Phys.} {\bf 322} 967-97
\item  Fu Y, Gui G, Liu Y and Qu C 2013 On the Cauchy problem for the integrable modified Camassa-Holm equation with cubic nonlinearity {\it J. Differ. Equ.} {\bf 255} 1905-38
\item  Liu T and Tian L 2013 Scattering problem for a modified Camassa-Holm equation {\it Int. J. Nonl. Sci.} {\bf 15} 178-81
\item  Sch\"afer T and Wayne C E 2004 Propagation of ultra-short optical pulses in cubic nonlinear media {\it Phys. D} {\bf 196} 90-105
\item  Matsuno Y 2007 Multiloop and multibreather solutions of the short pulse model equation {\it J. Phys. Soc. Japan} {\bf 76} 084003
\item  Matsuno Y 2008 Periodic solutions of the short pulse model equation {\it J. Math. Phys}. {\bf 49} 073508
\item  Matsuno Y 2009 Soliton and periodic solutions of the short pulse model equation 
       {\it Handbook of Solitons: Research, Technology and Applications} ed Lang S P and Bedore S H (New York Nova) Chapter 15  541-86
\item  Matsuno Y 2010 A direct method for solving the generalized sine-Gordon equation {\it J. Phys. A: Math. Theor.} {\bf 43} 105204
\item  Matsuno Y 2010 A direct method for solving the generalized sine-Gordon equation II {\it J. Phys. A: Math. Theor.} {\bf 43} 375201
\item  Matsuno Y 2013 Smooth multisoliton solutions and their peakon limit of Novikov's Camassa-Holm type equation with cubic nonlinearity {\it J. Phys. A: Math. Theor.} {\bf 46} 365203
\item  Ablowitz M J,  Kaup D J,  Newell A C and  Segur H 1974 The inverse scattering transform - Fourier analysis for nonlinear problems {\it Stud. Appl. Math.} {\bf 53}, 249-315
\item  Parker A 2004 On the Camassa-Holm equation and a direct method of solution. I. Bilinear form and solitary waves {\it Proc. R. Soc. Lond.} {\bf A 460} 2929-57
\item  Parker A 2005 On the Camassa-Holm equation and a direct method of solution. II. Soliton solutions {\it Proc. R. Soc. Lond.} {\bf A 461} 3611-32
\item  Parker A 2005 On the Camassa-Holm equation and a direct method of solution. III. N-soliton solutions {\it Proc. R. Soc. Lond.} {\bf A 461} 3893-911
\item  Matsuno Y 2005  Parametric representation for the multisoliton solution of the Camassa-Holm equation  {\it J. Phys. Soc. Japan}  {\bf 74} 1983-87
\item  Hirota R 1980 {\it Direct Methods in Soliton Theory}
       in {\it Solitons} ed RK Bullough and DJ Caudrey 
       Topics in Current Physics Vol. 17 (New York: Springer) p 157
\item  Matsuno Y 1984 {\it Bilinear Transformation Method} (New York: Academic Press)
\item  Hirota R and  Satsuma J 1976 $N$-soliton solutions of model equations for shallow 
       water waves  {\it J. Phys. Soc. Japan} {\bf 40} 611-2 
\item  Li J and Qiao Z 2013 Bifurcations and exact traveling wave solutions for a generalized Camassa-Holm equation {\it Int. J. Bifurcation and Chaos} {\bf 23} 1350057
\item  Parker A and Matsuno Y 2006 The peakon limits of soliton solutions of the Camassa-Holm equation {\it J. Phys. Soc. Japan}  {\bf 75} 124001
\item  Matsuno Y 2007 The peakon limit of the $N$-soliton solution of the Camassa-Holm equation {\it J. Phys. Soc. Japan}  {\bf 76} 034003
\item  Matsuno Y 2005 Multisoliton solutions of the Degasperis-Procesi equation and their peakon limit {\it Inverse Problems} {\bf 21} 1553-70
\item  Rabelo M L 1989   On equations which describe pseudospherical surfaces {\it Stud. Appl. Math.} {\bf 81} 221-48
\item  Leblond H and Mihalache D 2013 Models of few optical cycle solitons beyond the slowly varying envelope approximation {\it Phys. Rep.} {\bf 523} 61-126
\item  Matsuno Y 2006  Cusp and loop soliton solutions of short-wave models for the Camassa-Holm and Degasperis-Procesi equations
       {\it Phys. Lett. A} {\bf 359} 451-7
\item  Qiao Z and Xia B 2013 Integrable peakon systems with weak kink and kink-peakon interactional solutions {\it Front. Math. China} {\bf 8} 1185-96      
\item  Matsuno Y 2005 The $N$-soliton solution of the Degasperis-Procesi equation  {\it Inverse Problems} {\bf 21} 2085-101

\end{enumerate}

\end{document}